\documentstyle[prb,multicol,aps]{revtex}

\newcommand \be {\begin{equation}}
\newcommand \bea {\begin{eqnarray}}
\newcommand \ee {\end{equation}}
\newcommand \eea {\end{eqnarray}}
\newcommand \eps {\epsilon}
\newcommand \bi {\bibitem}
\newcommand \s {\sigma}

\makeatletter
\newcommand\erfc{\mathop{\operator@font erfc}\nolimits}
\makeatother
\input{epsf}
\begin{document}
\draft       

\title{Damage spreading transition in glasses: a probe for the ruggedness
of the configurational landscape}
\author{M. Heerema} 
\address{
Institute of Theoretical Physics\\ University of Amsterdam\\
Valckenierstraat 65\\ 1018 XE Amsterdam (The Netherlands).\\
E-Mail: heerema@phys.uva.nl}

\author{F. Ritort}
\address{Department of Fundamental Physics\\
Faculty of Physics\\ University of Barcelona\\
Diagonal 647\\ 08028 Barcelona (Spain).\\
E-Mail: ritort@ffn.ub.es}

\date{\today}
\maketitle

\begin{abstract}
We consider damage spreading transitions in the framework of
mode-coupling theory. This theory describes relaxation processes in
glasses in the mean-field approximation which are known to be
characterized by the presence of an exponentially large number of
meta-stable states. For systems evolving under identical but arbitrarily
correlated noises we demonstrate that there exists a critical
temperature $T_0$ which separates two different dynamical regimes
depending on whether damage spreads or not in the asymptotic long-time
limit. This transition exists for generic noise correlations such that
the zero damage solution is stable at high-temperatures being minimal
for maximal noise correlations. Although this dynamical transition
depends on the type of noise correlations we show that the asymptotic
damage has the good properties of an dynamical order parameter such as:
1) Independence on the initial damage; 2) Independence on the class of
initial condition and 3) Stability of the transition in the presence of
asymmetric interactions which violate detailed balance.  For maximally
correlated noises we suggest that damage spreading occurs due to the
presence of a divergent number of saddle points (as well as meta-stable
states) in the thermodynamic limit consequence of the ruggedness of the
free energy landscape which characterizes the glassy state.  These
results are then compared to extensive numerical simulations of a
mean-field glass model (the Bernasconi model) with Monte Carlo heat-bath
dynamics. The freedom of choosing arbitrary noise correlations for
Langevin dynamics makes damage spreading a interesting tool to probe the
ruggedness of the configurational landscape.
\end{abstract}



\section{Introduction}

The theoretical understanding of the dynamical behavior of glasses is a
long outstanding problem in statistical physics which has recently
revealed new aspects on the underlying mechanism responsible of the
glass transition \cite{SITGES,ANGELL,BCKM}.  The dynamical behavior of
glasses is characterized by the fast grow of the characteristic time of
relaxation processes in the vicinity of the glass temperature $T_g$.
This increase of the relaxation time, up to fifteen orders of magnitude
in a relatively small range of temperatures, is usually termed as the
viscosity anomaly. The first consideration of this anomaly, the
Vogel-Tamman-Fulcher law, goes back to the twenties.
However, there is still no satisfying and generally accepted
theoretical explanation for this singular behavior.

Currently, there are two main approaches to understand the glass
transition problem. One approach (the Adam-Gibbs-DiMarzio theory
\cite{AGM}) focuses on thermodynamic considerations and proposes the
existence of the ideal glass transition $T_g$. This is a singularity
where the configurational entropy of the under-cooled liquid vanishes at
$T_g$ and a second order phase transition characterized by a finite jump
in the specific heat occurs. This scenario has been rediscovered in the
framework of mean-field spin-glasses with one step of replica symmetry
breaking \cite{KITH}. In the mean-field approach, $T_g$ is the
temperature where configurational entropy vanishes (the so called
Kauzmann temperature) and also replica symmetry breaks.  In this paper
we will denote the glass transition both by $T_g$ and $T_s$ (in this
last case the subindex $s$ stands for statics). The other approach
relies on mode-coupling theory and describes the glass transition as a
strongly non-linear dynamical effect which induces long-term memory
properties in the correlation and response functions
\cite{GOTZE}. Consequence of these effects is the existence of a
dynamical singularity $T_d$ where ergodicity breaks and correlation
functions do not decay to zero. This dynamical transition $T_d$ is a
consequence of the mean-field character of the mode-coupling
approximation. Although these two approaches are apparently different
they have in common their mean-field character.

One of the most distinct features of glasses is the presence of a
complex free energy landscape.  The viscosity anomaly is a signature
of activated dynamics due to the existence of a rugged free energy
landscape with several maxima and minima separated by energy barriers
and saddle points which connect them.  One could think that the
existence of this type of landscape is a necessary ingredient to find
the previous scenario.  Well known results on the number of meta-stable
states in spin-glasses reveal that the interesting spin-glass
behavior emerges in systems with an exponential number of states
\cite{FOOT1}. For instance, models such as the spherical $p$-spin
interactions spin glass (with $p>2$), the Ising $p$-spin interactions
spin glass (with $p>3$), the Edwards-Anderson model and Potts glasses
are characterized by an exponentially large number of meta-stable
states. All these models are characterized by the presence of quenched
disorder which facilitates analytical treatments always at the level
of mean-field theory. In the absence of disorder, similar results are
found \cite{MPR,MPR2,BM} although exact calculations for the number of
meta-stable states turn out to be more difficult.  It is largely
believed that an exponentially number of meta-stable states is a
necessary condition for the existence of replica
symmetry breaking.

There have been also recent studies of exactly solvable models which, in
the absence of quenched disorder, also exhibit glassy behavior \cite{BG,BPR}
(and, in particular, activated behavior of the relaxation time). These
models are characterized by a small number of meta-stable states. What
makes them to display glassy behavior is the presence of entropy
barriers which leads to slow dynamics even in the absence of
meta-stability. Consequently, one is tempted to conclude that a rugged
free energy landscape with a large number of meta-stable states is not
essential to find glassy behavior but the presence of an enormous number
of flat directions in phase space.

A similar conclusion was reached in the study of the
Sherrington-Kirkpatrick spherical spin-glass \cite{SP2}. It was found
that an enormous number of zero modes are responsible for the slow
dynamics found in this model. Although that model does not have an
exponential large number of meta-stable states and does not show
activated dynamics it displays glassy behavior at low temperatures due to
the existence of flat directions around the meta-stable states
\cite{KL}.

A possible way to investigate the existence of flat directions in a
rugged free energy landscape is the study of the damage
spreading. Damage
spreading consists in the study of the dynamical evolution of the
distance $D(t)$ (to be defined later) between two systems
configurations evolving under the same dynamical rules and differing
only in their initial condition (the so called initial damage).

The study of the damage spreading problem (hereafter referred as DS) was
proposed by Kaufmann in the 60's for the study of the propagation of
mutations in the genotype in the biological growth of individuals
\cite{KA}. That is, how a small perturbation in the genotype
(microscopic level) manifests itself in the long-time term in the
fenotype (macroscopic level).  Since then, such a problem has received
considerable attention in the framework of statistical physics,
particularly in the middle 80's \cite{ALTRES}. Almost fifteen years ago
it was realized that DS could be a powerful tool to distinguish
different dynamical regimes in disordered systems, such as spin glasses
\cite{DW,DER}. Variants of damage spreading phenomena have also been
proposed to numerically investigate equilibrium correlation functions in
generic statistical systems \cite{CONIGLIO} and lattice gauge theories
\cite{PARISI,MARTIN}.  However, the initial enthusiasm and exciting
perspectives in the research of this problem decreased in subsequent
years after realizing that this transition was dependent on the type of
dynamics used. So the existence of the DS transition could have nothing
to do with the presence of a thermodynamic phase transition. Physicists
then started to systematically investigate the DS in well known ordered
systems such as the Ising model. In particular, much work has been
devoted to the study of the one-dimensional Domany-Kinzel automaton
\cite{DK} and the one-dimensional Ising model \cite{HR1,VT1}. 
The question of the
non-universality of damage spreading has been also emphasized in the
context of non-equilibrium phenomena such as domain growth by Graham,
Hernandez-Garcia and Grant \cite{GHG}.

So the question remains whether this transition has a truly physical
meaning or not. In this direction, Hinrichsen and Domany tried to give a
precise dynamic-independent definition for DS. To define a damage
spreading phase one must consider all possible dynamical procedures
which lead the system to thermal equilibrium. For discrete systems with
small number of nearest neighbors this definition can be implemented but
not in the general case (for instance, continuous systems) where an
infinity of dynamical rules can be always implemented.

The purpose of the present work is to present a detailed study of the DS
in a model with a rugged free energy landscape with an exponentially
large number of meta-stable states.  In particular, we will study the DS
in the $p$-spin spherical spin glass, an exactly solvable model for the
glass transition which is described by the Adam-Gibbs-DiMarzio scenario
and whose dynamics is described by the mode-coupling equations.  
To be more specific, we will study damage spreading for Langevin
dynamics, the simplest dynamics which is continuous in
time and satisfies ergodicity and detailed balance.
It must be stressed that although there are very few works on the DS problem using
this dynamics (Stariolo \cite{STA} and Graham et al. \cite{GHG}), the
majority of theoretical works in DS have considered discrete dynamics (in discrete systems).

We suggest that damage spreading can be used as a dynamical method to
evidenciate the existence of a large number of flat directions or saddle
points in phase space.  Also the existence of a dynamical transition
will be shown. Although we will check that damage spreading transitions
are strictly non-universal still it is possible to use the asymptotic
distance to define an order parameter for this dynamical transition. We
also anticipate that by considering correlations between the noises of
the two evolving systems an infinity of dynamical transitions can be
obtained. For Langevin dynamics, the case of maximally correlated noises
has a particularly interesting physical meaning.

Because our study considers the DS in the mode-coupling theory for
glasses it is expected to be generally valid for Langevin dynamics in
systems with a rugged free energy landscape such as realistic
glasses. Although the DS transition is non-universal and depends on
different dynamical rules (or cross-correlations between the stochastic
noises), we believe that this transition gives an interesting
information on the free energy landscape and could be investigated in
structural glasses. Being a signature of the existence of saddle-points
in phase space (i.e. points which separate stable and unstable phase
space directions) it is natural to expect that real glasses are good
systems to manifest these effects. This consideration makes our results
more attractive from the viewpoint of numerical simulations of realistic
glasses \cite{KOB}. Nevertheless we call the attention to the reader
that some of our claims in this paper are not generally proven (such as
the connections between damage spreading transitions and saddle points
in phase space) and the present reasarch should be seen as a first step
for a better understanding of  some of these questions.

The contents of the paper are divided as follows. The second section is
devoted to general considerations and definitions about the DS
problem. Section III demonstrates the
existence of a dynamical transition $T_0$ for DS. Section III is divided
in three subsections. Subsection III.A describes the mode-coupling
equations for the $p$-spin spherical spin glass, starting from a random initial configuration for different
correlations between the noises. This subsection also describes the
different numerical methods we have used to analyze the mode-coupling
equations for different cases. Subsection III.B analyzes the DS problem
starting from an equilibrium initial condition. Subsection III.C
analyzes the DS in the presence of asymmetric interactions. Section IV presents an
analysis of the damage spreading for glassy models with discrete
dynamics. Concretely, we study the DS transition in the Bernasconi model
with heat-bath dynamics. Finally section V presents
the conclusions. Two appendices are devoted to some technical issues.

\section{General considerations about DS}

In the most general framework the DS problem can be stated as
follows. Consider a dynamical system described by a generic variable
${\cal C}$ which denotes a given configuration. Suppose that the system
evolves under a deterministic dynamical rule $F$.  For sake of simplicity
we take a continuous time dynamics. The equation of motion reads,

\be
\dot{{\cal C}}_{t}=F({\cal C}_t)~~~~~~.
\label{eq1}
\ee

On top of the configuration ${\cal C}$ and the dynamical rule $F$ we
also need to define a distance in the phase space of configurations
(for instance, a Hamming distance for spin systems). This distance $D$
needs to satisfy the usual good properties, in particular
$D({\cal C}_t,{\cal C}_t)=0$ at all times. Suppose that we take two
initial configurations ${\cal C}_0,{\cal C'}_0$ with initial distance
$D_0=D({\cal C}_0,{\cal C'}_0)$ and consider the generic equal times
distance,

\be
D(t)= D({\cal C}_t,{\cal C'}_t)
\label{eq2}
\ee

where ${\cal C}(t)$ and ${\cal C'}(t)$ start from configurations ${\cal
C}_0$ and ${\cal C'}_0$ at time 0 and evolve under the {\it same dynamical
rule} $F$ eq.(\ref{eq1}).  Our main interest is to investigate the value of the
asymptotic long-time distance $D_{\infty}$,

\be
D_{\infty}=\lim_{t\to\infty} D(t)
\label{eq3}
\ee

Note that $D(t)=0$ if $D_0=0$.  Quite generally the
asymptotic distance $D_{\infty}$ will be a function of the type of
initial configurations ${\cal C}_0,{\cal C'}_0$ as well as their initial
distance $D_0$. The dependence of $D_{\infty}$ on those parameters is
governed by the dynamical properties of the deterministic rule, such as
chaotic properties and Lyapunov exponents.

One could extend this general problem to stochastic systems,
i.e. dynamical systems which evolve in the presence of a stochastic
noise. Let us consider two systems
described by the configuration variables ${\cal C}_t,{\cal C'}_t$ at time $t$ which
evolve following a Langevin dynamics,

\be
\dot{\cal C}_t=F({\cal C}_t)+\eta_t~~~~;~~~~\dot{\cal C'}_t=F({\cal C'}_t)+\eta'_t
\label{eq4}
\ee

where $F({\cal C})$ is a external force (which can eventually derive
from a potential although this is not a necessary condition) and
$\eta_t,\eta'_t$ are external white noises uncorrelated in time with
variance $2T$. Suppose now we make ${\cal C},{\cal C'}$ evolve following
eq.(\ref{eq4}) with the {\it same} realization of the stochastic noise
and starting from two different initial conditions. We are thinking of
noises statistically identical which coincide at equal times (i.e
$\eta_t=\eta'_t$) when ${\cal C}_t={\cal C'}_t$. So we choose
$\langle\eta_i(t)\eta_j(s)\rangle=\langle\eta'_i(t)\eta'_j(s)\rangle=2T
\delta(t-s)\delta_{ij}$ and cross-correlations
$\langle\eta_i(t)\eta'_j(s)\rangle=2T {\cal K}({\cal C}_t,{\cal
C'}_t)\delta(t-s)\delta_{ij}$ where ${\cal K}$ is a generic function
which satisfies the properties ${\cal K}({\cal C},{\cal C})=1$ and
$-1\le {\cal K}({\cal C},{\cal C'})\le 1,\,\forall {\cal C},{\cal C'}$.  In
the presence of stochastic noise, again $D(t)=0$ if $D(0)=0$. Note that
the role of the correlations is irrelevant for the evolution of the
independent systems ${\cal C},{\cal C'}$ but crucial for their
correlations and the equal times distance (\ref{eq2}).  Different
choices of the cross-correlation ${\cal K}$ for Langevin dynamics are
the analogous of different dynamical rules in discrete dynamics such as
Monte Carlo (these rules could be Glauber, Metropolis or heat-bath among
others). 

Now we are interested in the asymptotic long-time distance
eq.(\ref{eq3}). Quite generally,
$D_{\infty}$ will be a function of the type of initial condition (for
instance, random or stationary), the initial distance $D_0$, the
intensity of the noise $T$, and the cross-correlator ${\cal K}$.
For the case in which $\eta_t=\eta'_t$ (${\cal K}=1$), we will find that
there is a dynamical phase transition at a finite temperature $T_0$
below which the asymptotic distance is different from zero.  The origin
of this dynamical transition can be explained in plain physical
words. In eq.(\ref{eq4}) there is competition between two different
terms. On the one hand, the {\em force} term $F({\cal C}_t)$ propagates
the error (or damage) in the initial configuration. Instead, the noise
$\eta_t$ acts in the same way in both systems smearing out possible
differences in the initial condition. In other words, the stochastic
noise is the synchronizing force which tries to make both evolving
configurations to merge in time while the force term amplifies the
initial damage playing the role of a {\em noise}.  This
argument only applies if ${\cal K}=1$. In the general case $-1\le {\cal
K}\le 1$ the noise does not necessarily synchronize both systems and
its effect is similar to that of the {\em force}.  In this case, the
asymptotic distance will be also a function of the cross-correlation
${\cal K}$ (i.e. the dynamical rule).

To understand better the role of the cross correlation ${\cal K}$ let us
consider as a starting point the simple problem of a particle which moves inside an
harmonic potential $V(x)=\frac{1}{2}x^2$ following a Langevin dynamics,

\be \frac{\partial x(t)}{\partial t}=-\frac{\partial V(x)}{\partial
x}+\eta(t)
\label{eq5}
\ee
 
where $\eta(t)$ is a stochastic white noise of variance $2T$. The
configuration $C$ corresponds to the position $x$ of the particle and we
define a distance between two configurations $x,y$ as $D(x,y)=(x-y)^2$.
Take now two identical particles $x,y$ and make them follow
eq.(\ref{eq5}) both with the same stochastic noises $\eta,\eta'$ and
cross-correlation ${\cal K}(x,y)$. For simplicity we will take a $x,y$
symmetric cross-correlation ${\cal K}={\cal K}(D)$.  If $<..>$ stands
for average over dynamical histories then the distance
$D(t)=<(x(t)-y(t))^2>$ satisfies the following equation,

\be
\frac{\partial D(t)}{\partial t}=-2D(t)+4T(1-{\cal K}(D))~~~~~~.
\label{eq6}
\ee

This equation has several stationary solutions depending on ${\cal
K}$. Obviously $D=0$ is a stationary solution (remember ${\cal K}(0)=1$)
which implies ${\cal K'}(0)<0$. Only if $-1/2T<{\cal K}'(0)<0$ the
solution $D=0$ is stable. On the other hand, for ${\cal K}'(0)< -1/2T$
the solution $D=0$ is unstable and $D(t)$ converges to another
stationary solution (which can be shown it is always stable). So, there
is a dynamical transition at $T_c=1/(2|{\cal K'}(0)|)$ where the
asymptotic distance $D_{\infty}$ changes from zero ($T<T_c$) to
$D_{\infty}=D^*$ where $D^*$ satisfies the identity $D^*=2T(1-{\cal
K}(D^*))$. The asymptotic distance is then given by,
$D_{\infty}=\lim_{t\to\infty} D(t)=D^*\Theta(T^*-T)$ independent of the
value of the initial distance $D_0$ between the two particles.  So when
${\cal K'}(0)$ does not vanish, already for the simple harmonic
oscillator there is more than one stationary solution. In other words,
the effect of the cross-correlation term ${\cal K}(D)$ manifests through
the appearance of more than one stationary solution. As the reader can
imagine the discussion turns out to be more difficult for other more
complex potentials.

Although this dependence of the asymptotic distance $D_{\infty}$ on the
cross-correlation ${\cal K}$ is an intrinsic property of damage
spreading it does not necessarily imply that this kind of dynamical
phase transitions do not give any relevant information on the physical
properties of the system.  What this really means is that the results
concerning DS may depend on particular forms of cross-correlation
function between the noises (similarly as happens for discrete dynamics
where different dynamical rules yield different results). Nevertheless,
generic results for DS may be obtained for correlators which satisfy
quite general conditions (as we will see later).

The major part of work in this paper will be devoted to the
study of maximal cross-correlations, i.e ${\cal K}=1$ although the
results are extrapolable to more generic cross-correlators.
For Langevin dynamics, the case ${\cal K}=1$ is particularly appealing
for two reasons.  

On the one hand, it followed from the simple example of the harmonic oscillator
that ${\cal K}=1$ implies ${\cal K'}=0$ everywhere. 
Below we will argue that this observation holds also for
more complex potentials.
More specific, we argue that in the case ${\cal K}=1$ there is a single stationary
solution $D_{\infty}=0$ for any finite system and for any confining
potential which diverges in the boundaries (i.e $V(x)\to \infty$ when
$x\to\pm \infty$). Clearly, the harmonic oscillator is a trivial
case where the asymptotic distance always goes to zero independently
of the temperature $T$ and how far is $x$ from $y$ at $t=0$. In order
to justify our assertion let us take a more complex potential of two wells separated by a finite
barrier, for instance a particle moving inside an harmonic plus a
quartic term potential $V(x)=-\frac{1}{2}x^2+\frac{\lambda}{4}
x^4$. In this case the potential has two wells located at $x=\pm
\frac{1}{\sqrt{\lambda}}$. If two systems described by the variables
$x,y$ start to evolve within the same well (i.e. $x,y>0$ or $x,y<0$)
they will always tend to finish in the same final configuration
because the synchronizing effect of the noise is, at very long times,
the dominant effect. If they start in different wells the conclusion
is also the same because there is always a finite probability that a
strong fluctuation in the stochastic noise drives both particles in
the same well. This conclusion, which holds for maximal
cross-correlations ${\cal K}=1$, can be generalized for any potential
with a {\em finite} number of wells separated by {\em finite} energy
barriers. There is always a finite probability that a fluctuation of
the noise can take both particles into the same well and hereafter the
distance $D(t)$ between both configurations would tend exponentially
fast to zero. Obviously this argument applies only for finite
barriers, finite wells as well as finite temperature. At zero
temperature the synchronizing effect of the stochastic noise is absent
and the asymptotic distance may not vanish and show a non-trivial
dependence on the initial distance.

From the discussion above it follows that the role of the
thermodynamic limit $N\to\infty$ in the DS is also crucial. In this limit the height of the barriers
or the number of wells (i.e. meta-stable states) may diverge. The first
case happens, for instance, in the ferromagnetic Ising model where the
time-reversal symmetry of the Hamiltonian is broken below $T_c$. 
The second case is realized in spin-glass models where the number of
meta-stable states is exponentially large with $N$. In both cases, a
finite fluctuation of the stochastic noise (even with ${\cal K}=1$) may not synchronize
the system and $D_{\infty}$ can be a non trivial function of both the
temperature and the initial distance. To be more precise $D_{\infty}$ is
defined as follows,

\be
D_{\infty}=\lim_{t\to\infty}\lim_{N\to\infty} D(t)
\label{eq7}
\ee

Note that for $N$ finite we expect $\lim_{t\to\infty} D(t)=0$ so it is
crucial that the thermodynamic limit $N\to\infty$ is taken before the
infinite-time limit. Taking the limits in the reverse order will result
in that $D_{\infty}$ will always vanish at finite temperature. 
Note that this discussion applies only when the $D=0$ stationary solution is
stable. This is indeed satisfied for ${\cal K}=1$ but may be also
fulfilled in more general situations with ${\cal K}<1$ and $D=0$ still
being an stationary stable solution.

There is another property of the case ${\cal K}=1$ which makes it
particularly interesting. Up to now our discussion was limited to
different mathematical properties of damage spreading transitions. But,
what about their physical significance? Suppose we take two generic
statistical systems described by the set of variables $\lbrace
x_i,y_i\rbrace, i=1,..,N$ which evolve under the Langevin dynamics,

\be
\frac{\partial x_i}{\partial t}=F_i(\lbrace x(t)\rbrace)+\eta_t~~~;~~~
\frac{\partial y_i}{\partial t}=F_i(\lbrace
y(t)\rbrace)+\eta'_t\label{new}\ee

where, as before, $\eta,\eta'$ are white noises with cross-correlation
${\cal K}$ which we will suppose is a generic function of the Hamming
distance $D$. Let us suppose that the force derives from a potential
$F_i(\lbrace x\rbrace)=-\frac{\partial V(\lbrace x\rbrace)}{\partial
x_i}$. If we define the new variables $z_i=x_i-y_i$ we may obtain,
subtracting both equations in (\ref{new}),

\be
\frac{\partial z_i}{\partial t}=F_i(\lbrace y(t)+z(t)\rbrace)-
F_i(\lbrace y(t)\rbrace)+\nu_i(t)
\label{new2}\ee

where $\nu$ is a stochastic white noise of zero mean and variance
$4T(1-{\cal K}(x,y))$. The solution $z_i=0$ is a stationary solution of
(\ref{new2}). A linear stability analysis around that solution yields
the equation,

\be
\frac{\partial z_i}{\partial t}=H_{ij}(\lbrace y(t)\rbrace)z_j(t)+\nu_i(t)
\label{new3}
\ee

where $H_{ij}(y)=-\frac{\partial^2V(\lbrace y\rbrace)}{\partial
y_i\partial y_j}$ is the Hessian matrix evaluated at the point
$(y_1,..,y_N)$.  The solution $z_i=0$, corresponding to vanishing
Hamming distance $D=0$, is stable if the Hessian is negative
definite. The presence of the stochastic noise $\nu$ in (\ref{new3})
decreases the stability of the $D=0$ solution. Because ${\cal K}\le 1$
we conclude that the region where the $D=0$ stationary solution is
maximally stable corresponds to the case when ${\cal K}=1$ because $\nu$
vanishes. So ${\cal K}=1$ is the cross-correlation for which the
stationary solution $D=0$ is maximally stable.  For ${\cal K}=1$
equation (\ref{new3}) is quite appealing and evidenciates the physical
origin of the DS transition. An instability of the $D=0$ solution may
appear when an eigenvalue of the Hessian matrix $H_{ij}$ vanishes. This
corresponds to a saddle point of the potential landscape $V(\lbrace
y\rbrace)$. Due to ergodicity, the $x,y$ systems sample all the possible
configurations. So the asymptotic distance $D_{\infty}$ is a direct
measure of the stability of the $D=0$ solution along all possible
configurations (weighted with their corresponding statistical Boltzmann
weight). In other words, a instability in the solution $D_{\infty}=0$
and the existence of a DS transition is indication of the presence of
saddle points in the potential landscape of systems $x,y$. If ${\cal
K}<1$ the temperature of the DS transition will depend on the particular
form of ${\cal K}$ (actually we will see later, in the study of
mode-coupling equations, that it depends through the value ${\cal
K'}(D=0)$). Furthermore, due to the destabilizing effect of the noise
the damage spreading transition will increase when ${\cal K}$ decreases
so ${\cal K}=1$ yields the lowest damage spreading transition
temperature among all possible cross-correlations ${\cal K}(D)$ for
which the $D=0$ solution is stable.  In the presence of fixed points for
the asymptotic distance other than $D=0$ the stationary solution
$D_{\infty}=0$ may become unstable because the noise $\nu$ is too strong
(similarly to what happens in the harmonic oscillator example) and damage
spreading does not evidenciate anymore the existence of saddle points.
In other words, saddle points may be {\em observed} only by studying
the $T$ dependence of the basin of attraction of the $Q=0$ stationary
solution supposing it stable at very high-temperatures (free case).
Remark that a similar argument has already been presented by Loreto, Serva and
Vulpiani \cite{LSV} for systems described by a single variable $x(t)$ in
a potential field $V(x)$.  

We have argued above that for Langevin dynamics the maximal cross-correlator
${\cal K}=1$ is a special case, resulting in a simplification of the
problem.
We emphasize that in other dynamical systems it is unclear whether or not the
maximal cross-correlator ${\cal K}=1$ plays the same role in the
context of damage spreading.
This is due to the complexity of cross-correlations.  

In the next section, we will analyze in detail the DS dynamics in the
mode-coupling theory of glasses. As has been previously said these
equations describe the relaxation processes and dynamics in glasses
(in the under-cooled liquid regime) in the mean-field approximation
and represent the dynamical behavior of systems with an exponentially
large number of meta-stable states.

\section{Damage spreading in Mode-Coupling theory}

Mode-coupling theory describes relaxational processes in glasses. In
the nutshell, mode-coupling theory corresponds to an exact re-summation
of an infinite series of diagrams in the hydrodynamic theories.  The
kind of diagrams which the mode-coupling approximation selects are
those which precisely survive in the mean-field limit of some
realistic models.  So, a way to get mode-coupling equations is by
considering exact dynamical theories for mean-field disordered
spin-glass models \cite{KITH}. Spherical spins allow for an exact closure of
the dynamical equations in terms of correlation and response functions
(as was shown by Crisanti, Horner and Sommers \cite{CHS} in the
$p$-interactions spherical spin glass), leading to many analytical
results.  Although spherical spins are unrealistic
(compared to Ising spins) they capture the essentials of the dynamics
which is universally found in a large variety of models.
Whereas for $p=2$ the
physical description of the model is quite simple \cite{SP2} the
behavior turns out to be much more interesting for $p>2$ where an
exponentially large number of meta-stable states are present
\cite{CS2}. In this type of models meta-stability plays a very
important role so, according to the arguments of the previous section,
we expect to get interesting results for the DS
transition. Forthcoming subsections analyze this transition in
detail. A good review on the main results obtained in this model have
been collected and reported by Barrat \cite{BARRAT}.

\subsection{Random initial configuration}

This section is devoted to the study of the DS problem in the
mode-coupling equations. This section describes some preliminary work
already presented in \cite{HR} but here we present a more extended
research of the problem to include asymmetry, different class of
initial conditions as well as general cross-correlations of the
noises.  The simplest solvable model whose dynamics is described by
the off-equilibrium mode-coupling equations is the spherical $p$-spin
glass model introduced by Crisanti, Horner and Sommers \cite{CHS}. In
this case, the configurations are described by $N$ continuous spin
variables $\lbrace\s_i; 1\le i \le N\rbrace$ which satisfy the
spherical global constraint $\sum_{i=1}^N\s_i^2=N$. 

The Langevin dynamics of the model is given by,

\be
\frac{\partial \s_i}{\partial t}=F_i(\lbrace\s\rbrace)-\mu\s_i+\eta_i
\label{eqa1}
\ee

\noindent
where $F_i$ is the force acting on the spin $\s_i$ due to the
interaction with the rest of the spins,

\be
F_i=\frac{1}{(p-1)!}
\sum_{(i_2,i_3,...,i_p)}\,J_i^{i_2,i_3,..,i_p}\s_{i_2}\s_{i_3}..\s_{i_p}~~~.
\label{eqa2}
\ee

The term $\mu$ in eq.(\ref{eqa2}) is a Lagrange multiplier which ensures
that the spherical constraint is satisfied at all times and the noise
$\eta$ satisfies the fluctuation-dissipation relation
$\langle\eta_i(t)\eta_j(s)\rangle=2T \delta(t-s)\delta_{ij}$ where
$\langle...\rangle$ denotes the noise average. The
$J_i^{i_2,i_3,..,i_p}$ are quenched random variables with zero mean and
variance $p!/(2N^{p-1})$. The interactions $J_i^{i_2,i_3,..,i_p}$
are symmetric under the interchange of the super-indices $i_2,i_3,..,i_p$
but in the most general case may not be symmetric under the exchange of
the subindex $i$ with a generic super-index. So, for instance
$J_i^{i_2,i_3,..,i_p}\ne J_{i_2}^{i,i_3,..,i_p}$. Most of the studies
undertaken in this model concentrate on the symmetric case where $J_i^{i_2,i_3,..,i_p}$
is symmetric under the permutation of all possible indices.  This case
is particularly interesting because there exists an energy function such
that the force $F_i$ derives from a Hamiltonian or potential function
$ F_i=-\frac{\partial {\cal H}}{\partial \s_i} $
so there exists an stationary state described by a
Boltzmann-Gibbs distribution. Due to the
mean-field character of the model, the dynamical equations depend on the
statistical properties of the force only through its correlations.
On the other hand, the statistical properties of the force $F_i$ depend
on the correlations of the $J's$. The simplest case \cite{CKLP} corresponds
to correlations of the type,

\be
\overline{J_{i_1}^{i_2,i_3,..,i_p}\,J_{i_k}^{i_1,..,i_{k-1},i_{k+1},..,i_p}}=
\alpha\frac{p!}{2N^{p-1}}
\label{eqcorrJ}
\ee

for every $k$. So if $\alpha=1$ we recover the symmetric case while for $\alpha=0$ we
obtain the asymmetric case. Eq.(\ref{eqcorrJ}) implies the following
statistical properties for the force $F_i$ \cite{CKLP},

\be
\overline{F_i(\lbrace \s\rbrace) F_j(\lbrace
\tau\rbrace)}=\delta_{ij}f'(q)+(1-\delta_{ij}) \alpha f''(q) \frac{\tau_i\s_j}{N} 
\label{eqcorrF}
\ee 

where $f(q)=q^p/2$. In the asymmetric case $\alpha=0$ the forces are
completely uncorrelated at different sites. Hence the set of equations
(\ref{eqa1}) become uncorrelated and the problem can be partially
solved. This particular case will be analyzed later.  For $\alpha<1$ there does not
exist an energy function ${\cal H}$ which drives the system to thermal
equilibrium and the fluctuation-dissipation theorem is not fulfilled.

We define the overlap between two configurations of the spins
$\s,\tau$ by the relation $Q=\frac{1}{N}\sum_{i=1}^N\s_i\tau_i$ so
the Hamming distance between these two configurations is,

\be
D=\frac{1-Q}{2}
\label{eqa3}
\ee

\noindent
in such a way that identical configurations have zero distance and
opposite configurations have maximal distance $D=1$. Then we consider
two copies of the system $\lbrace\s_i,\tau_i\rbrace$ which evolve under
the same statistical noise (\ref{eqa1}) with cross-correlation ${\cal
K}$ but with different initial conditions.  We assume the
cross-correlator to be a function of the Hamming distance $D$ or the
overlap $Q$.  As a consequence, other choices for the cross-correlator
(in general this could depend on both configurations ${\cal C},{\cal
C'}$) may change our results obtained below for ${\cal K}(Q) \neq 1$.
The major part, however, is concerned with ${\cal K}=1$ (we will explain
why) and will not suffer from this.  In this section we restrict our
attention to random initial configurations (i.e equilibrium
configurations at infinite temperature) with initial overlap $Q(0)$. The
case of initial equilibrium configurations will be analyzed in the next
subsection. The different set of correlation functions which describe
the dynamics of the system are given by,

\bea
C(t,s)=(1/N)\sum_{i=1}^N\langle\s_i(t)\s_i(s)\rangle=(1/N)\sum_{i=1}^N\langle\tau_i(t)\tau_i(s)\rangle
\label{eqa4a}\\
R(t,s)=(1/N)\sum_{i=1}^N\frac{\partial\langle\s_i\rangle}{\partial h^{\s}_i}=
(1/N)\sum_{i=1}^N\frac{\partial\langle\tau_i\rangle}{\partial h^{\tau}_i}\label{eqa4b}\\
Q(t,s)=(1/N)\sum_{i=1}^N\langle\s_i(t)\tau_i(s)\rangle\label{eqa4c}
\eea

\noindent
where $<..>$ denotes the average over dynamical histories and
$h^{\s}_i,h^{\tau}_i$ are fields coupled to the spins $\s_i,\tau_i$
respectively. In what follows we take the convention $t>s$. The previous
correlation functions satisfy the boundary conditions, $C(t,t)=1,
R(s,t)=0, \lim_{t\to (s)^+} R(t,s)=1$ while the two replica overlap
$Q(t,s)$ defines the equal time overlap $Q_d(t)=Q(t,t)$ which yields the
Hamming distance at equal times or damage $D(t)$ through the relation
(\ref{eqa3}). Following standard functional methods \cite{CUKU,BARRAT}
it is possible to write a closed set of equations for the previous
correlation functions. Some details of the computation are shown in
Appendix A. The final result is:

\bea
\frac{\partial C(t,s)}{\partial t}+\mu(t)C(t,s)=
\frac{p}{2}\int_0^s du R(s,u)C^{p-1}(t,u)+\alpha\frac{p(p-1)}{2}
\int_0^t du R(t,u)C(s,u)C^{p-2}(t,u)\label{eqa5a}\\
\frac{\partial R(t,s)}{\partial t}+\mu(t)R(t,s)=\delta(t-s)
+\alpha\frac{p(p-1)}{2}
\int_s^t du R(t,u)R(u,s)C^{p-2}(t,u)\label{eqa5b}\\
\frac{\partial Q(t,s)}{\partial t}+\mu(t)Q(t,s)=
\frac{p}{2}\int_0^s du R(s,u)Q^{p-1}(t,u)
+\alpha\frac{p(p-1)}{2}
\int_0^t du R(t,u)Q(u,s)C^{p-2}(t,u)\label{eqa5c}
\eea

while the Lagrange multiplier $\mu(t)$ and the diagonal correlation
function $Q_d(t)$ obey the equations,

\bea
\mu(t)=T+\frac{p(1+\alpha(p-1))}{2}\int_0^tdu R(t,u)C^{p-1}(t,u)\label{eqa6a}\\
\frac{1}{2}\frac{\partial Q_d(t)}{\partial t}+\mu(t)Q_d(t)=T{\cal K}(Q_d(t))+
\frac{p}{2}\int_0^t du R(t,u)Q^{p-1}(t,u)
+\alpha\frac{p(p-1)}{2}
\int_0^t du R(t,u)Q(t,u)C^{p-2}(t,u)\label{eqa6b}~~~~~~.
\eea

Note that the cross-correlation ${\cal K}(Q_d)$ only enters
explicitly through the equation (\ref{eqa6b}) so it does not affect the
evolution of a single replica. The whole set of equations is
quite involved. For the correlation $C$ and response functions $R$
eqs.(\ref{eqa5a},\ref{eqa5b},\ref{eqa6a}) several results are known,
in particular their behavior in the equilibrium regime (where
time-translational invariance is satisfied and the
fluctuation-dissipation theorem is obeyed) as well as in the
non-stationary aging regime \cite{CUKU}.

In what follows we analyze different dynamical fixed points of
eq.(\ref{eqa6b}) and show the existence of a dynamical instability in
the DS equations.

\subsection{Fixed-point analysis for a generic cross-correlation ${\cal K}$}

\begin{figure}
\begin{center}
\leavevmode
\epsfysize=150pt{\epsffile{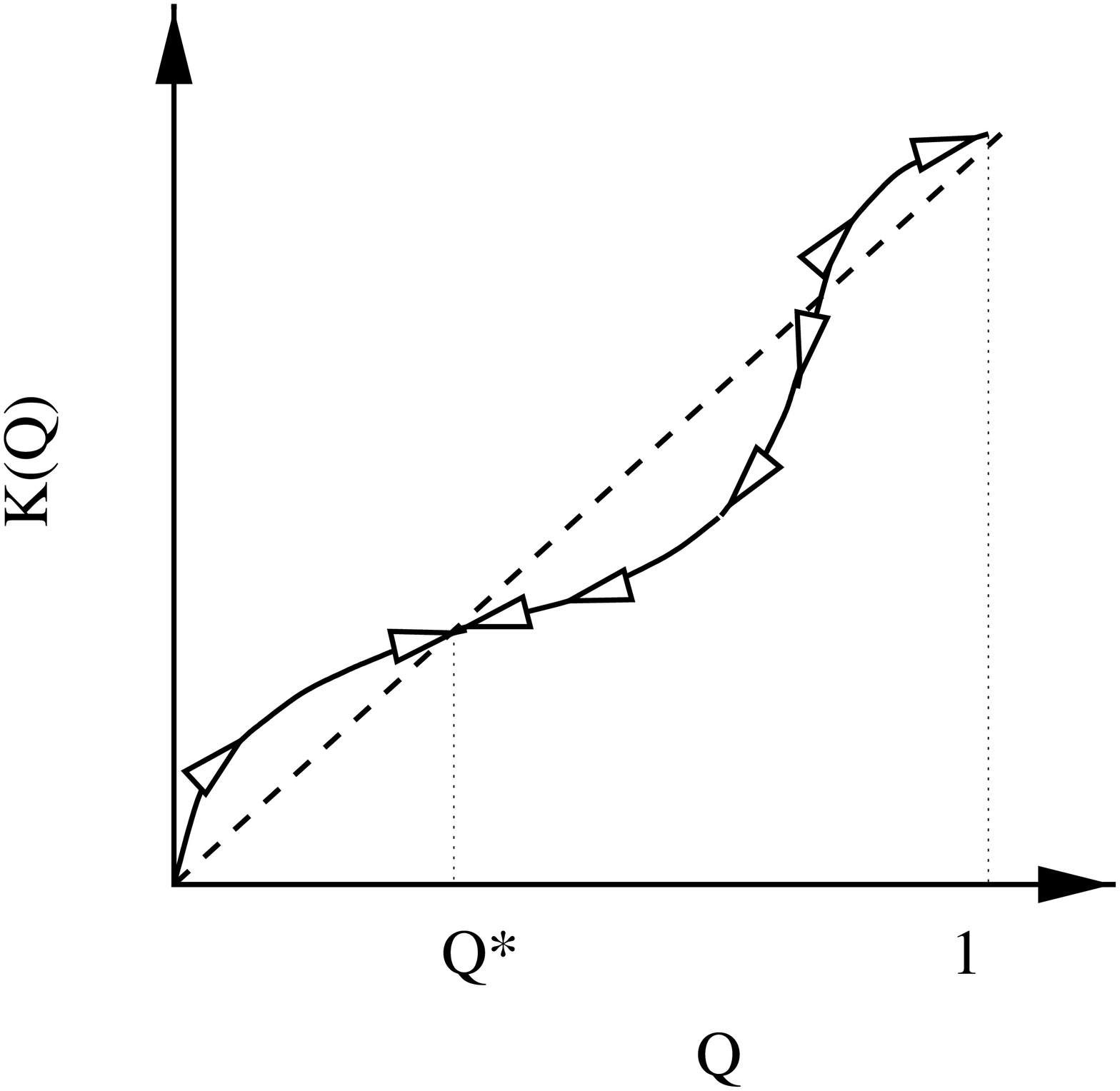}}
\end{center}
  \protect\caption[1]{Flow diagrams and fixed points for a generic
  correlator ${\cal K}$ at infinite temperature. $Q^*$ and $1$ are
  stable fixed points and $0$ and the fixed point between $Q^*$ and 1
  are unstable.
  \protect\label{FIXED1} }
\end{figure}

\begin{figure}
\begin{center}
\leavevmode
\epsfysize=150pt{\epsffile{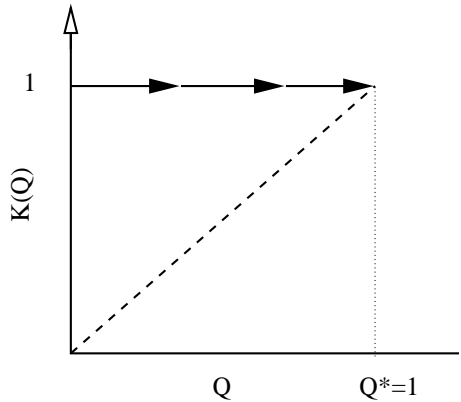}}
\end{center}
  \protect\caption[1]{Flow diagram for ${\cal K}=1$.
  \protect\label{FIXED2} }
\end{figure}

Different type of dynamical regimes may be distinguished depending on
the cross-correlator ${\cal K}$. Our analysis is similar to that
performed in section II for the simple harmonic oscillator. Different
fixed points for the dynamical equations can be analyzed from equation
(\ref{eqa6b}). If the temperature $T$ is very large then eq.(\ref{eqa6b}) becomes,

\be
\frac{1}{2}\frac{\partial Q_d(t)}{\partial t}=T({\cal K}(Q_d(t))-Q_d(t))
\label{K1}
\ee

where we have used $\mu=T$ using eq.(\ref{eqa6a}). Equation (\ref{K1})
can be exactly solved. The stationary solutions are given by ${\cal
K}(Q)=Q$. In figure~(\ref{FIXED1}) we analyze the different solutions
for a generic ${\cal K}$. We find that there are different stationary
solutions corresponding to all possible intersections of the two curves
($Q$ and ${\cal K}(Q)$). A linear stability analysis of (\ref{K1})
reveals that stationary solutions $Q^*$ are stable if ${\cal
K'}(Q^*)<1$. A dynamical flow diagram can be constructed where the
region of stability is indicated by different arrows. Regions where
${\cal K}(Q)>Q$ satisfy $\frac{\partial Q_d}{\partial t}>0$ and regions
where ${\cal K}(Q)<Q$ satisfy $\frac{\partial Q_d}{\partial t}<0$. So in
this case one may depict a diagram of all possible high-temperature
dynamical phases which separate regions with different fixed-point
attractors. Damage spreading transitions will strongly depend on the
type of cross-correlator. The case ${\cal K}=1$ is shown in
figure~(\ref{FIXED2}) where there is a unique attractor at $Q=1$ at very
high-temperatures. This analysis of the different dynamical phases is
valid only at very high temperatures. As soon as the temperature is
finite and starts to decrease some of the stable fixed points may become
unstable and other unstable points may become stable. The damage
spreading transition correspond to the appearance of an instability in
one of these high-temperature fixed points. As we will see below, the
damage spreading transition temperature may be different for different
fixed-points since it depends on the value of ${\cal K'}(Q^*)$ which may
vary for different fixed-points $Q^*$.

In what follows most of our discussion will concentrate on the
particularly interesting case ${\cal K}=1$ which has a unique fixed
point at $Q^*=1$. Although the analysis may be extended to other
fixed-points this case is also the most interesting according to our
preceding discussion in section II. As we will check below, this case
also defines the lowest damage spreading temperature $T_0$ among all the
possible cross-correlators ${\cal K}$ for which the fixed-point $Q^*=1$
is stable.

\subsubsection{Existence of $T_0$ : lower and upper bound for $\alpha=1$}

A first glance to equations (\ref{eqa5c}),(\ref{eqa6b}) reveals that the
overlap $Q(t,s)$ and its diagonal part $Q_d(t)$ are coupled each other
through the correlation $C(t,s)$ and response function $R(t,s)$.  The
trivial solution $Q(t,s)=C(t,s)$ and $Q_d(t)=1$ corresponds to the case
where the initial conditions are the same, $Q_d(0)=1$ and the distance
$D(t)=0$ for all times. This high-temperature fixed-point (hereafter we
will denote it by HT) corresponds to $D_{\infty}=0$ and is
asymptotically reached by the dynamics for high enough temperatures
under certain conditions of the cross-correlator ${\cal K}$ (see the
preceding discussion). In what follows we concentrate our attention in
the case $\alpha=1$ where there is a stationary solution for $C,R$ and
$\mu$ corresponding to the equilibrium measure. Numerical integration of
those equations (see later) reveals that the typical time needed to
reach that solution grows if temperature decreases. At a given
temperature (which we identify with $T_0$) there is an instability in
the dynamical equations (\ref{eqa5c}),(\ref{eqa6b}) and the asymptotic
solution differs from the HT one. We did not succeed in finding an
explicit expression for $T_0$ but we have been able to show its
existence and find a lower and upper bound for its value (see Appendix
B): 

\be
\sqrt{\frac{p-2}{2(1-{\cal K'}(1))}} \le T_0\le \sqrt{\frac{p}{2(1-{\cal K'}(1))}}
\label{equpper}
\ee

Note that for the particular case $p=2$ and ${\cal K'}(1)=0$ the
inequality (\ref{equpper}) yields $T_{0}\le 1$. Taking into account that
(\ref{equpper}) was derived under the assumption $T_{0}\ge T_s=1$
(i.e. we supposed we were in the high-temperature regime) this is not
inconsistent with the result $T_{0}=1$ derived by Stariolo \cite{STA}.

Note that both the lower and upper bound for $T_0$ diverge when ${\cal
K'}(1)=1$. This limiting value sets a condition on the possible
cross-correlations ${\cal K}(Q)$ where the $Q=1$ solution is
stable. {\em Only} those functions ${\cal K}(Q)$ for which ${\cal
K'}(1)<1$ are those for which $Q=1$ is linearly stable at very
high-temperatures. According to our discussion in section II the
appearance of DS in this case is related to the presence of a divergent
number of saddle points which mark the onset of a dynamical
instability. Because ${\cal K'}(1)<0$ then (according to
(\ref{equpper})) the limiting case ${\cal K'}(1)=0$ (for instance, if
${\cal K}(Q)=1$) sets the lowest value of the temperature $T_0$ where DS
appears.  This is important because it means, that whatever correlator
${\cal K}$ we consider (such that the solution $Q=1$ is stable for
high-enough temperatures) damage spreads below
$T_0=\sqrt{\frac{p-2}{2}}$.
Remark that in the general case $p\ge 3$ the dynamical instability
temperature $T_0$ stays well above any relevant critical temperature
($T_s$ or $T_d$).
 
In the next section we discuss the behavior of the asymptotic distance
as a function of temperature. For simplicity, our analysis is restricted
to the case ${\cal K}=1$ for which most of the numerical work has been
done. We will see that $D_{\infty}$, for a given specification of the
correlator ${\cal K}$, seems indeed to play the role of a dynamical
order parameter in DS transitions.

\subsubsection{Numerical analysis}

In general it is too complicated to obtain an analytical solution of
the set of equations (\ref{eqa5a})--(\ref{eqa6b}). We shall devote
this section to a numerical study of the equations
(\ref{eqa5a})--(\ref{eqa6b}) for the DS problem.  Although in
some particular cases an exact solution can be found (see below) this is not
the general situation.

\begin{figure}
\begin{center}
\leavevmode
\epsfysize=160pt{\epsffile{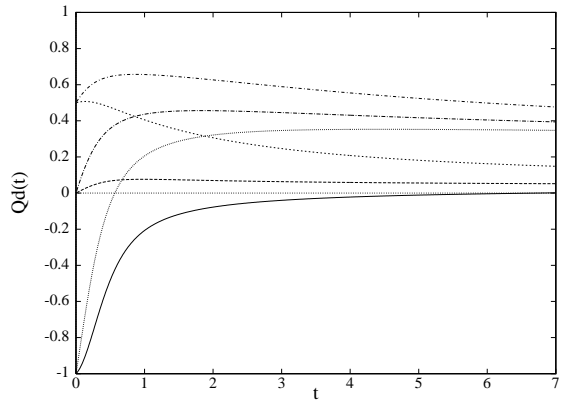}}
\end{center}
  \protect\caption[1]{$Q_d(t)$ for $p=3$ ($\alpha=1,{\cal K}=1$) at
  temperatures $T=0.1,0.5$ (from bottom to top at large times) for three
  different values of the initial overlap $Q_d(0)=-1,0,0.5$ as a
  function of time. The continuous lines are the numerical integrations
  with time step $\Delta t=0.01$.  \protect\label{FIG1B} }
\end{figure}

First, one could investigate the long-time limit of $Q_d$ via a
numerical integration of the set of equations
(\ref{eqa5a})--(\ref{eqa6b}).  However, the CPU time and the memory
needed to do this grows very fast with time because of the integrals
occurring in the equations.  Thus the spreading of damage at large times
can only be obtained from the dynamical equations doing some
extrapolations.  This enlarges the error in the estimate of
$D_{\infty}$, especially in cases where $Q_d(t)$ is a non-monotonic
function of time.  In figure~(\ref{FIG1B}) we
show how the overlap $Q_d(t)$ depends on the initial condition. Although
direct extrapolations from numerical data of the value of the asymptotic
damage are difficult, the figure is not incompatible with an
independence of $D_{\infty}$ on the initial condition. Another more
powerful technique is necessary to corroborate this result.

An alternative approach to obtain the long-time behavior of time
dependent variables with high accuracy was introduced by Franz, Marinari
and Parisi\cite{FMP} to study the long-time behavior of the energy.
Here we extend their method to analyze the asymptotic behavior of
$D(t)$. In their method they first decompose the time dependent
variables in a series expansion before extrapolating for large times
with the help of Pade approximants.  For the DS problem, it leads to a
Taylor expansion of the correlation function $C$, the response function
$R$ and the overlap $Q$:

\be
\label{series-c,r,q}
C(t,s)=\sum_{k=0}^{\infty} \sum_{l=0}^{\infty} c_{kl} t^k s^l \, ,
\qquad R(t,s) = \sum_{k=0}^{\infty} \sum_{l=0}^{\infty} r_{kl} t^k s^l
\, , \qquad Q(t,s) = \sum_{k=0}^{\infty} \sum_{l=0}^{\infty} q_{kl}
t^k s^l \, , \ee where $c_{00}=r_{00}=1$ since $C(t,t)=1$ and
$\lim_{t\to (s)^+} R(t,s)=1$ \cite{FOOT3}. Moreover, $\mu(t)$ and the diagonal correlation
function $Q_d(t)$ can be written as:
\be
\label{series-mu,dam}
\mu(t) = \sum_{k=0}^{\infty} \mu_k t^k \, , \qquad Q_d(t) =
\sum_{k=0}^{\infty} \sum_{l=0}^k q_{(k-l)l} t^k \, , \ee with
$\mu_0=T$ and $q_{00}$ is a parameter identical to the value of
$Q_d$ at $t=0$.  Assuming always $t>s$, the dynamical equations
(\ref{eqa5a})--(\ref{eqa6b}) can be transformed into recurrence
relations for the coefficients of the expansion.  To this end one
first substitutes (\ref{series-c,r,q}) and (\ref{series-mu,dam}) into
(\ref{eqa5a})--(\ref{eqa6b}) under the constraint $t>s$, then
calculate the integrals and finally rearrange terms.

Numerically, the coefficients of the expansion are now readily obtained.
In case $p=3$, the first $80$ coefficients of the expansions can be
computed on a RISC workstation in a few hours. 
However, for larger values of $p$ the computational effort is larger.  

To ensure high accuracy of the asymptotic extrapolation, one needs a
large radius of convergence of the series expansion.  A good method to
enlarge the radius of convergence of a series expansion is using Pade
approximants.  In this method one introduces two polynomials $U_m(t)$
and $V_k(t)$ of degree at most $m$ and $k$, respectively.  The goal is
to choose $U_m(t)$ and $V_k(t)$ for given $m$ and $k$ such that
$Q_d(t)$ and $U_m(t)/V_k(t)$ are equal at $t=0$ and have as many
equal derivatives as possible at $t=0$.

The computations have been performed for the symmetric case $\alpha=1$
and ${\cal K}=1$.  Moreover, three different initial conditions have
been considered: a) anti-correlated random initial conditions with
$Q_d(0)=-1$, b) uncorrelated random initial conditions with $Q_d(0)=0$
and c) partially correlated random initial conditions with $Q_d(0)=0.5$.

To check that the extrapolations $D_{\infty}$ using the Pade approximants
are correct, the Pade series have been compared with numerical integrations
of the dynamical equations.
Indeed, the Pade series and the numerical integration fit closely \cite{HR}.

%

\begin{figure}
\begin{center}
\leavevmode
\epsfysize=180pt{\epsffile{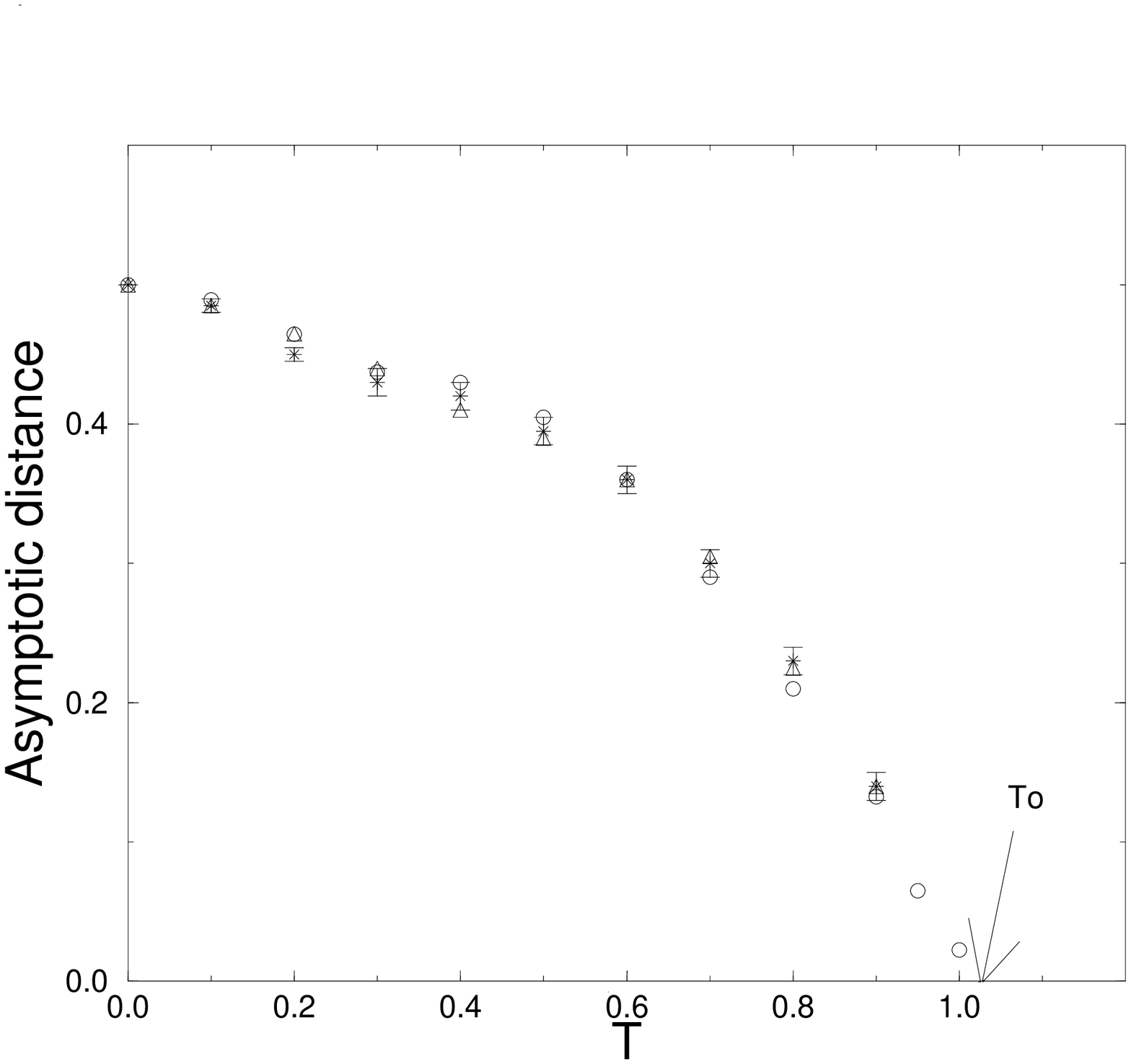}}
\end{center}
  \protect\caption[2]{Asymptotic distance $D_{\infty}$ for $p=3$
($\alpha=1,{\cal K}=1$) obtained from the Pade analysis of the series
expansions for different initial conditions $D_0=1$ (circles), $D_0=0.5$
(triangles), $D_0=0.25$ (stars). Typical error bars are shown for the
last case.  \protect\label{FIG2} }
\end{figure}

\begin{figure}
\begin{center}
\leavevmode
\epsfysize=150pt{\epsffile{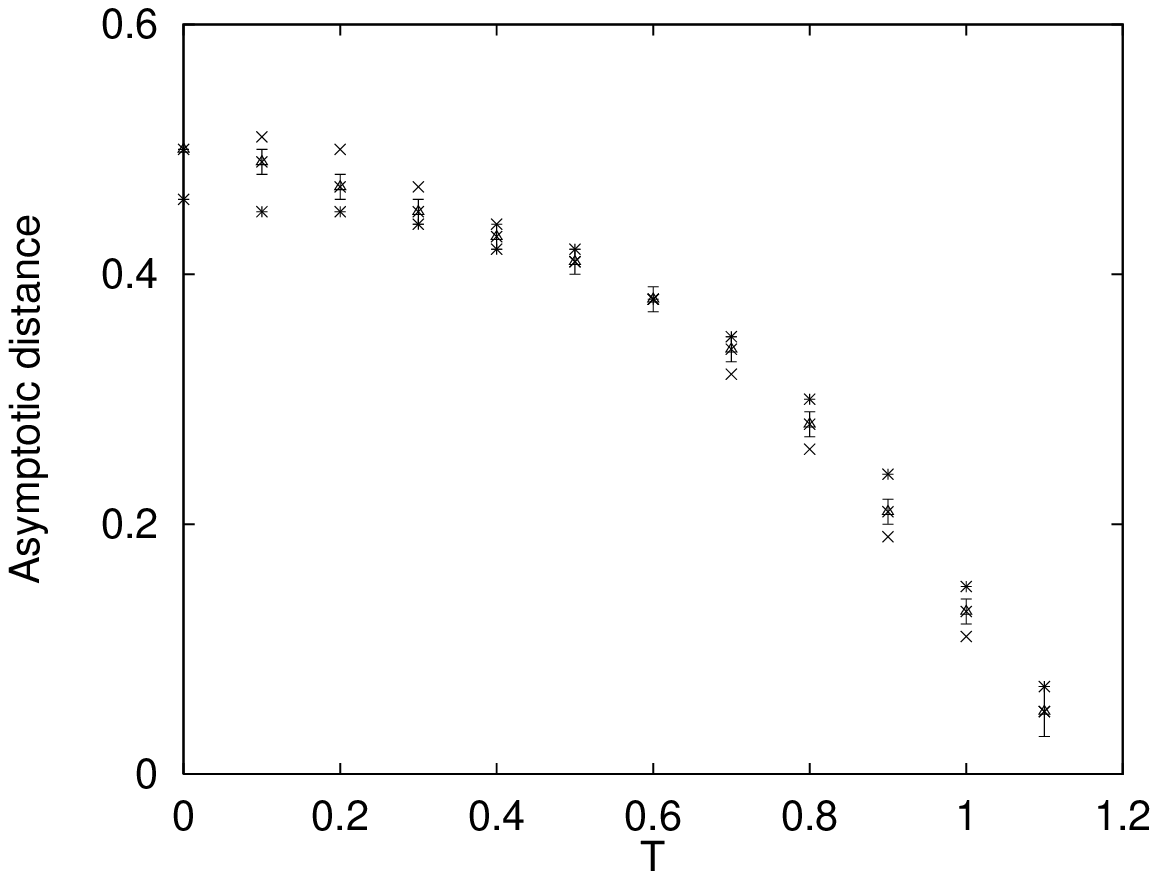}}
\end{center}
\protect\caption[3]{Asymptotic distance $D_{\infty}$ for $p=4$
($\alpha=1,{\cal K}=1$) obtained from the Pade analysis of the series
expansions for different initial conditions $D_0=1$ (crosses), $D_0=0.5$
(triangles), $D_0=0.25$ (stars). Typical error bars are shown for the
case $D_0=0.5$.  \protect\label{FIGp4} }
\end{figure}

The estimate for $D_{\infty}$ is obtained by division of the highest
order coefficients of $P_m(t)$ and $Q_k(t)$, i.e., by $a_m/b_k$.
Moreover, an asymptotic estimate can be obtained assuming a power law
decay of the equal time overlap: $Q_d(t)=Q_d(\infty)+At^{-\gamma}$.
The analysis of $D_{\infty}$ suffers in some cases from a small radius
of convergence (even with Pade) as well as from the presence of poles
in the Pade expansion.  The results are displayed for $p=3$ in
figure~\ref{FIG2} and for $p=4$ in figure~\ref{FIGp4} for cases a), b)
and c) as a function of the temperature.  Let us remark that a lower
number of coefficients in the Taylor expansion in the case $p=4$ with
respect to $p=3$ leads to a less accurate estimate of the asymptotic
distance. 

Inspection of figures~\ref{FIG2} and \ref{FIGp4} reveals that the
dynamical transition $T_0$ is in the predicted regime eq. (\ref{equpper}).
It can be estimated more accurately from the relaxation time
$\tau_{\rm relax}$ associated to the decay of the distance $D(t)$ to zero.
Starting from high temperatures, we assume the relaxation time to
diverge at $T_0$ according to a power law: $\tau_{\rm relax} \simeq
(T-T_0)^{-\gamma}$. 
We thus have found $T_0(p=3)=1.04 \pm 0.02$ with $\gamma \simeq 1.1 \pm
0.1$ and $T_0(p=4)=1.13 \pm 0.02$ with $\gamma \simeq 1.1 \pm 0.1$.
 
We conclude that for all temperatures, both $p=3$ and $p=4$, the
asymptotic distance is independent of the initial distance.  This is in
contrast with the case $p=2$, where dependence of the initial distance
is found for the low temperature region \cite{STA}. We must say that we
have obtained the same results, as in cases $p=3,4$, for a model which
is a combination of the $p=2$ and $p=4$ spherical spin-glass model
\cite{THEO}. For a certain range of parameters, this model is known to
have a continuous phase transition with continuous replica symmetry
breaking and without collapse of the configurational entropy.  So the
first-order character of the spin-glass transition found in our model
for $p>3$ is not essential for the appearance of the DS
transition. Still, that model \cite{THEO}, is also characterized by the
presence of an exponentially large number of meta-stable states. From the
point of view of the form of the dynamical equations, the fact that
$T_0$ is present for $p>2$ as well as in a model which combines $p=2$
and $p=4$ is a consequence of the non-linearity in the coupling between
the damage $Q_d(t)$ and the two-time correlation function $Q(t,s)$ which
occurs for all $p>2$. From the physical point of view, this independence
of $D_{\infty}$ on the initial distance is quite appealing.  In general,
one would expect that $p=4$ is quite similar to $p=2$ due to the
presence of the time-reversal symmetry. The fact that the damage does
not feel this symmetry for $p>2$ means that the separation of dynamic
trajectories does not occur in the borders or maxima which separate
equilibrium states but within saddle points of the phase space which
divide configurations separated by finite energy barriers.  This is
supported by the fact that the transition occurs at a temperature much
higher than $T_s$ and, as we will see in the following section, from the
fact that it happens starting already from an equilibrium configuration.
The asymptotic value $D_{\infty}$ can, on the basis of our computations,
be regarded as an order parameter for the transition at $T_0$. Although
$D_{\infty}$ and the transition $T_0$ itself do depend on the specific
choice of the correlator ${\cal K}$ it is interesting that the
asymptotic state does not depend on the initial distance. For a better
understanding of the physical origin of this transition we shall
consider the case of equilibrium initial conditions in the next section.

\subsection{Equilibrium initial condition}

We have seen in the previous section that the asymptotic distance
$D_{\infty}$ is a non-trivial function of the temperature which is
finite below $T_0$ and vanishes above $T_0$. The relevance of the
existence of the meta-stable states has been already pointed out in
previous sections specially when the cross-correlator is maximal ${\cal
K}=1$. The fact that the DS transition exists suggests that the nature
of this phase transition is related to the corrugated properties of the
free energy landscape. To check this result it is convenient to
investigate the DS transition starting from an equilibrium
configuration. In this case the system starts from a stationary state
and remains there forever. At high temperatures this state is the
paramagnetic so in this case the DS is a direct check of the ruggedness
of the paramagnetic state.  In fact, we will find that if we start from
an initial equilibrium condition \cite{FOOT2}, then the DS transition
persists and actually coincides with the previous $T_0$ found for a
random initial configuration. This reinforces the idea of $D_{\infty}$
as a dynamical order parameter for the DS transition for a given choice
of the correlator ${\cal K}$. Again, for sake of simplicity, we restrict
our analysis here to the case ${\cal K}=1$.

The analysis of the dynamical equations for an equilibrium initial
condition follows the same steps as for the random case but now we
must impose a Gibbs distribution for the configurations $\s$ and
$\tau$ at time 0. Nevertheless, there is a point which must now be
considered. Let us take $T>T_s$ (i.e we will suppose equilibrium
configurations in the paramagnetic phase). Suppose that $p$ is odd and
we take an equilibrium configuration at temperature $T$. To impose
$Q_0=-1$ or $D_0=1$ we must take $\s_i=-\tau_i$ for all $i$. Because
the energy is an odd function of the spin variables we have $E(\lbrace
\s \rbrace)=-E(\lbrace \tau \rbrace)$. If the equilibrium energy is
not zero (this happens everywhere except at $\beta=0$) we cannot put
both configurations at equilibrium at the same temperature (because
the temperature uniquely determines the value of the equilibrium
energy). Then, if both initial conditions $\sigma$ and $\tau$ are
equilibrium initial configurations their overlap $Q(0)$ must
vanish. Actually, for $T>T_s$ two equilibrium configurations do have
overlap zero with probability 1 and overlap different from zero with
probability exponentially small with $N$. So if we take the
thermodynamic limit before the infinite time limit it is clear that we
must start with zero initial overlap. To be more precise, the
probability that two equilibrium configurations $\lbrace \s
\rbrace,\lbrace \tau \rbrace$ in the paramagnetic phase do have
overlap $q=\frac{1}{N} \sum_{i=1}^N\s_i\tau_i$ is given by,

\be
P(q)\sim exp(-N\beta f(q))
\label{eqcost}
\ee

where $f(q)$ is the free energy cost to find a correlation $q$ between
the configurations. Clearly, because $q=0$ corresponds to the
equilibrium value in the paramagnetic phase, $f(q)$ has a minimum at
$q=0$ so we can write $f(q)=constant+q^2/(2\beta\chi_{SG})$,

\be
P(q)\sim  exp(-\frac{Nq^2}{2\chi_{SG}})
\label{eqcost2}
\ee

with $\chi_{SG}=N<q^2>$ is the spin-glass susceptibility. Above $T_s$
the $\chi_{SG}$ is finite and the probability to have $q\ne 0$ is
exponentially small with $N$. Now the cost in free energy $f(q)$ has
two parts, a cost in energy $u(q)$ and a cost in entropy $s(q)=\beta
(u(q)-f(q))$. The cost in energy vanishes at infinite temperature and
the full cost of $f(q)$ is due to the entropy. So only at infinite
temperature (i.e random initial configurations, the case considered in
preceding sections) we can impose an initial condition with initial
non-zero overlap.

The equivalent of the dynamical equations
eqs.(\ref{eqa5a},\ref{eqa5b},\ref{eqa5c}) can be easily obtained for
$T>T_s$ in the replica symmetric approximation \cite{BARRAT}. The correlation function
$C(t,s)$ and the response function $R(t,s)$ are time-translational
invariant and satisfy the fluctuation-dissipation theorem
$TR(t)=-\frac{\partial C(t)}{\partial t}\Theta(t)$. The $C(t)$
satisfies the equation,

\be
\frac{\partial C(t)}{\partial t}+TC(t)+\frac{\beta p}{2}\int_0^t\,du 
C^{p-1}(t-u) \frac{\partial C(u)}{\partial u}=0
\label{eqC}
\ee

with $C(0)=1$. The two-times overlap satisfies the equation,

\bea
\frac{\partial  Q(t,s)}{\partial t}+(T+\frac{p\beta}{2})Q(t,s)-
\frac{p\beta}{2}\int_0^{s}\, du Q^{p-1}(t,u)\frac{\partial
C(s-u)}{\partial u}-\frac{p\beta}{2}\int_0^{t}\, du Q(s,u)\frac{\partial
C^{p-1}(t-u)}{\partial u}=0
\label{eqQt1t2}
\eea
with the initial condition $Q(t,0)=Q(0,t)=0$. The diagonal part,
$Q_d(t)=Q(t,t)$ is given by,

\bea
\frac{1}{2}\frac{\partial  Q_d(t)}{\partial
t}+(T+\frac{p\beta}{2})Q_d(t)\nonumber\\
-T-\frac{p\beta}{2}\int_0^{t}\, du Q^{p-1}(t,u)\frac{\partial
C(t-u)}{\partial u}-\frac{p\beta}{2}\int_0^{t}\, du Q(t,u)\frac{\partial
C^{p-1}(t-u)}{\partial u}=0\label{eqQtt}
\eea

with the initial condition $Q_d(0)=0$. Now we are in equilibrium so
$\mu(t)=T+\frac{p\beta}{2}$ \cite{BARRAT}.  We have looked for a
time-translational invariant solution for $Q(t,s)$ (i.e a solution of
the type $Q(t,s)=Q_d(s)\hat{Q}(t-s)$ for $t>s$) but we have not
found it (even for $p=2$). Our numerical results suggest that such an asymptotic
solution does not exist.

\begin{figure}
\begin{center}
\leavevmode
\epsfysize=150pt{\epsffile{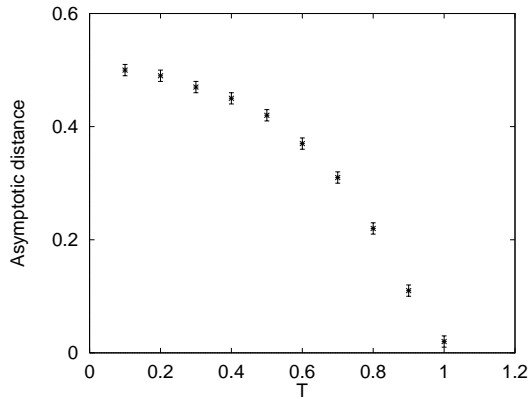}}
\end{center}
  \protect\caption[4]{Asymptotic distance $D_{\infty}$ as a function of
    the temperature, starting from an equilibrium configuration, for
    $p=3$ ($\alpha=1,{\cal K}=1$).  \protect\label{FIGeq} }
\end{figure}

Using as before a series expansion in the time-dependent variables and
Pade approximants, we have estimated the asymptotic distance for
equilibrium initial conditions, i.e., for $Q(0)=0$.
The results are displayed in figure~\ref{FIGeq} for different temperatures.
The divergence of the relaxation time leads to $T_0(p=3)=1.01 \pm
0.04$ with $\gamma=1.4 \pm 0.3$ which indicates that $T_0$ coincides
with the result obtained for random initial conditions.  
This supports the idea that the transition at $T_0$ is of a dynamical
nature and unrelated to the existence of a thermodynamic phase transition.

\subsection{The non-symmetric $\alpha\ne 1$ case}

As we saw in the previous section, one of the most interesting results
concerning the DS transition is the fact that $D({\infty})$ is a
non-trivial quantity which does not depend on the initial condition (and
depends only on the cross correlation ${\cal K}$).  In equilibrium
thermodynamics, this is one of the features of order parameters which
separate different equilibrium phases. In the stationary state, when
fluctuation-dissipation theorem is obeyed, the order parameter is a a
quantity which characterizes the equilibrium state and (in the absence
of ergodicity breaking) does not depend on the initial condition.


In order to present a convincing proof on this result we have
investigated the general non-symmetric case $\alpha\ne 1$. A difficulty
inherent to the extrapolations made from figures 1 and 2 is the fact
that, below $T_0$, the convergence of the distance $D(t)$ towards its
asymptotic value $D_{\infty}$ is very slow (a power law in
time). Consequently, both numerically or using the Pade method it is
very difficult to extrapolate to the asymptotic value. As the asymmetry
of the interactions eq.(\ref{eqcorrJ}) is turned on (i.e if $\alpha <1$)
the relaxation of the system to the stationary state turns out to be
faster. Actually for the asymmetric case ($\alpha=0$) or the
antisymmetric case ($\alpha=-\frac{1}{p-1}$, see later) the relaxation
of the distance $D(t)$ is nearly exponential. We have no reasons to
suppose that the independence of the asymptotic value on the initial
distance is $\alpha$ dependent. Our analysis for $\alpha<1$ suggests
that the independence of $D(t)$ on the initial distance $D_0$ holds for
all generic values of $\alpha$. 


It is important to note that, for $\alpha<1$, there is no equilibrium
stationary state and the fluctuation-dissipation theorem is not
satisfied.  Still we expect, for $p>2$, the DS transition to survive
for any quantity of finite asymmetry. The reason is that the DS transition
(even for $\alpha=1$) is inherently a non-equilibrium transition so the
effect of the asymmetry may not change the character of that transition.

The non-symmetric case for $p=2$ has been already considered by
Crisanti and Sompolinsky ten years ago \cite{CS}. By
assuming that, in the stationary state, the correlation and response
functions are time-translational invariant they succeeded in showing
that the thermodynamic transition $T_s=1$ for $\alpha=1$ turned out to
be unstable against the asymmetry for any value of $\alpha< 1$. They
also derived the result $\mu(\infty)=\sqrt{1+T^2}$ for
$\alpha=0$. 

In the next paragraphs we study the cases $\alpha=0,-\frac{1}{p-1}$
starting from a random initial configuration in some detail. Unless
stated we will consider the case ${\cal K}=1$.

\subsubsection{The asymmetric case $\alpha=0$}

The case $\alpha=0$ is quite interesting. The equation for the response
function (\ref{eqa5b}) simplifies considerably,

\be
\frac{\partial R(t,s)}{\partial t}+\mu(t)R(t,s)=\delta(t-s)
\label{eqas1}
\ee

which can be readily written, using $R(t,s)=\frac{z(s)}{z(t)}$ with $z(t)=exp\bigl ( \int_0^t \mu(t')dt' \bigr )$. The equation for
the correlation function becomes, 

\be
\frac{\partial C(t,s)}{\partial t}+\mu(t)C(t,s)=
\frac{p}{2}\int_0^s du \frac{z(u)}{z(s)}\,C^{p-1}(t,u)
\label{eqas3}
\ee

Define the new function $A(t,s)=z(t)C(t,s)z(s)$. In terms of this new
function eq.(\ref{eqas3}) is,

\be
\frac{\partial A(t,s)}{\partial t}=\frac{p}{2}\int_0^s du 
\frac{A^{p-1}(t,u)}{z^{p-2}(u)z^{p-2}(t)}~~~~~~~.
\label{eqas4}
\ee

From eq.(\ref{eqa6a}) it is easy to derive a equation for $z(t)$,

\be \frac{1}{2}\frac{\partial z^2(t)}{\partial t}=
Tz^2\,+\,\frac{p}{2}\int_0^tdu
\frac{A^{p-1}(t,u)}{z^{p-2}(u)z^{p-2}(t)}~~~~~~~.
\label{eqas5}
\ee

Equations eq.(\ref{eqas4}),(\ref{eqas5}) form a closed set of equations
which can be solved with the initial conditions $A(t,t)=z^2(t)$,
$z(0)=1$. Once this set of equations is solved one can also find a
solution for the overlap $Q(t,s)$ in eq.(\ref{eqa5c}). Again, we define
$B(t,s)=z(t)Q(t,s)z(s)$ which satisfies the equation,

\be
\frac{\partial B(t,s)}{\partial t}=
\frac{p}{2}\int_0^s du \frac{B^{p-1}(t,u)}{z^{p-2}(u)z^{p-2}(t)}\label{eqas6}
\ee

and the equal-times overlap eq.(\ref{eqa6b}) $b(t)=B(t,t)$ satisfies the equation,

\be
\frac{1}{2}\frac{\partial b(t)}{\partial t}=Tz^2(t)+
\frac{p}{2}\int_0^t du \frac{B^{p-1}(t,u)}{z^{p-2}(u)z^{p-2}(t)}\label{eqas7}
\ee

with $b(0)=Q_d(0)$. Note that this set of equations is quite involved for
$p>2$. Only for $p=2$ they dramatically simplify (the case considered by
Crisanti and Sompolinsky) and become linear. For general $p$ the previous
equations are non-linear. We have not succeeded in finding the asymptotic
solution of these equations although we have guessed the results from
the numerical results. We find that the DS transition is still present
at finite temperature for $p>2$. The analytical expression for $T_c$ is
given by 

\be
T_c=\sqrt{\frac{p-2}{2}}
\label{eqas8}
\ee

Note that $T_c$ coincides precisely with the lower bound previously
derived in eq.(\ref{equpper}). The asymptotic value of $\mu(t)$ is given
by $\mu(\infty)=\sqrt{1+T^2}$ and is $p$-independent. The asymptotic
distance for $p=3,4$ is given by \cite{FOOT4},

\be
D_{\infty}=\frac{1-T-(1-T_c)\bigl (\frac{T}{T_c}\bigr )^2}{2}
\label{eqas9}
\ee

\begin{figure}
\begin{center}
\leavevmode
\epsfysize=160pt{\epsffile{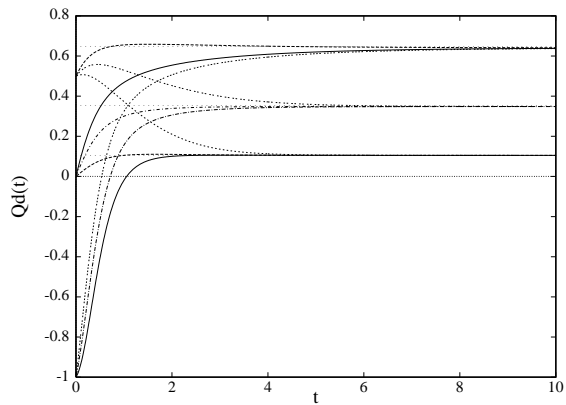}}
\end{center}
  \protect\caption[1]{$Q_d(t)$ ($p=3, \alpha=0, {\cal K}=1$) with
  $Q_d(0)=0,0.5,-1$ as a function of time for temperatures
  $T=0.1,0.3,0.5$ (from bottom to top at large times). The horizontal
  dotted lines are the theoretical guesses.\protect\label{FIGASA} }
\end{figure}

\begin{figure}
\begin{center}
\leavevmode
\epsfysize=160pt{\epsffile{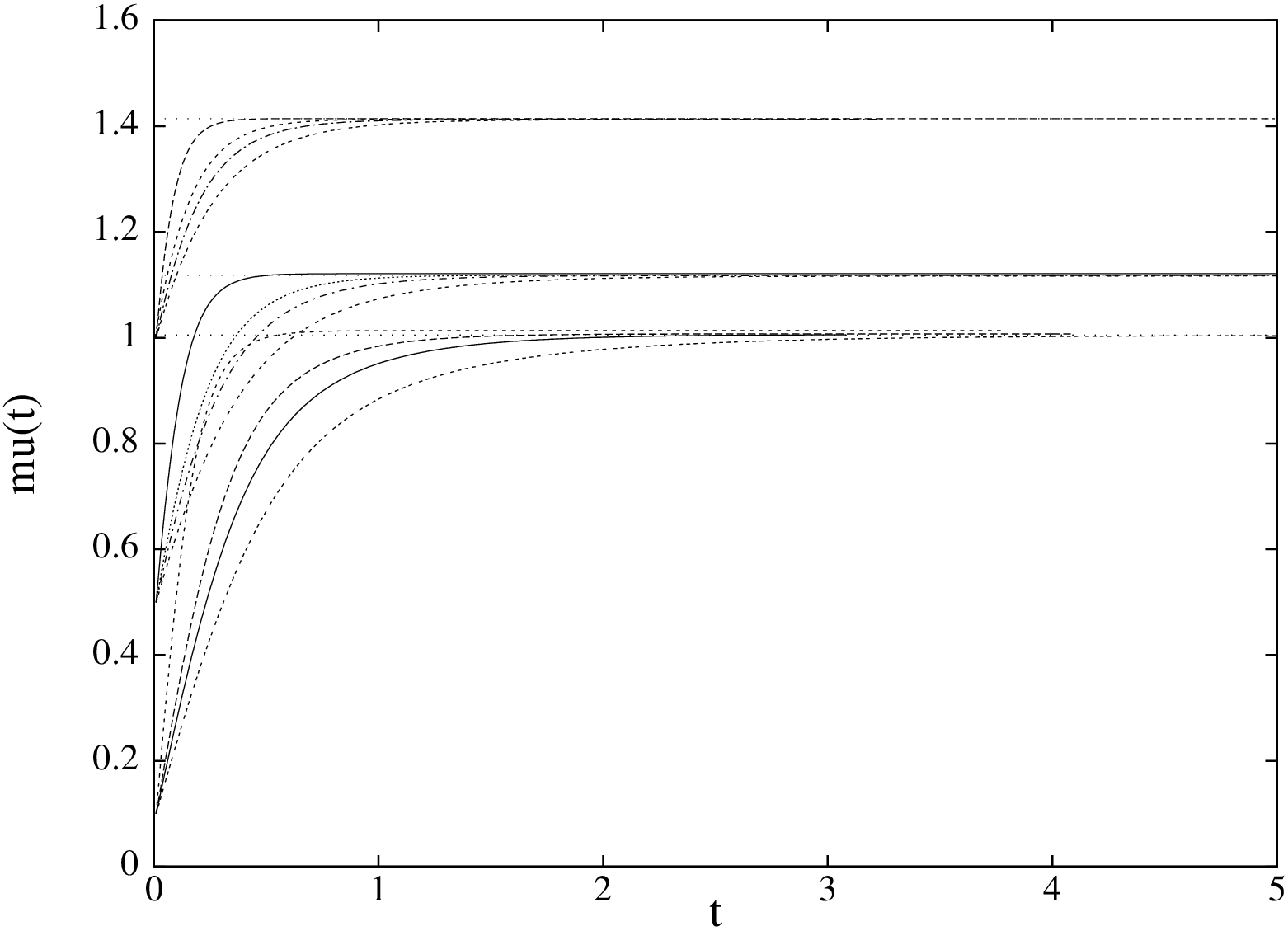}}
\end{center}
  \protect\caption[1]{$\mu(t)$ for $\alpha=0,{\cal K}=1$ ($p=3$ solid,
  $p=4$ dashed) with $Q_d(0)=-1$ as a function of time for temperatures
  $T=0.1,0.3,0.5,0.7$ (from bottom to top). The horizontal line is the
  theoretical prediction. Note that $\mu(0)=T$. \protect\label{FIGASB} }
\end{figure}

A full theoretical derivation of this results remains an interesting
open problem. In what follows we compare our results obtained by numerical
integrations with time step $\Delta t=0.01$ with these theoretical
guesses.  In figure~\ref{FIGASA} we show the overlap $Q_d(t)$ as a
function of time for different temperatures below $T_c$ for
$p=3$. Note that the asymptotic value clearly does not depend on the
initial condition. The horizontal dotted lines correspond to the
asymptotic value eq.(\ref{eqas9}).  This figure unambiguously
demonstrates that the asymptotic distance does not depend on the value
of the initial overlap $Q_d(0)$.  In figure~\ref{FIGASB} we show
$\mu(t)$ for $p=3,4$ compared with the theoretical prediction
$\mu(\infty)=\sqrt{1+T^2}$.

\subsubsection{The antisymmetric case $\alpha=-\frac{1}{p-1}$}

This is an extremal case where
$\overline{J_{i_1}^{i_2,i_3,...,i_p}J_{i_k}^{i_1,..,i_{k-1},i_{k+1},..,i_p}}$is
maximally negative. Physically this means that the force felt by an
spin $i$ due to a multiplet ${\cal M}$ of $p-1$ spins is as much
opposite as possible as the force which feels another spin contained in
that multiplet due to the action of another multiplet ${\cal M'}$ of
$p-1$ spins constructed from the rest of the $p-2$ spins in the previous
multiplet ${\cal M}$ plus the spin $i$.  In the particular case $p=2$
this can be easily achieved making $J_i^j=-J_i^j$ which corresponds,
according to eq. (\ref{eqcorrJ}), to $\alpha=-1$.  But in general,
$\alpha$ can never be equal to $-1$ for $p>2$. Take $p=3$ and three
couplings $J_i^{jk},J_j^{ik},J_k^{ij}$.  It is clear that if
$J_i^{jk}=-J_j^{ik}$ and $J_i^{jk}=-J_k^{ij}$ then $J_j^{ik}\ne
-J_k^{ij}$. It can be easily shown that the minimum value for $\alpha$
is given by $\alpha=-\frac{1}{p-1}$.

Interestingly, this case can be exactly solved for the correlation and
response function. Although it turns out to be quite difficult to solve
for the overlap function we will analyze here a general correlator ${\cal
K}$. For $\alpha=-\frac{1}{(p-1)}$ eqs. (\ref{eqa5a}-\ref{eqa6b})
considerably simplify because $\mu(t)$ in eq.(\ref{eqa6a}) is time
independent. Because the initial configuration was taken random at time
0 this means that the stationary state follows completely random
configurations.

To solve the equations define the following correlators,

\be
C(t,s)=c(t,s)\exp{(-T(|t-s|))};~~~R(t,s)=r(t,s)\exp{(-T(t-s))}~~~t\ge s~~~
\label{eqas10}
\ee

In this case the dynamical equations for $C$ and $R$ simplify. A particular solution
for $c(t,s)$ and $r(t,s)$ can be found which simplifies dramatically the
dynamical equations. This solution is given by $c(t,s)=r(t,s)=f(t-s)$.
This time-translational invariant solution is consistent with ${\em all}$
dynamical equations for ${\em all}$ times. The final closed equation for $f(t)$
is given by,

\be
\frac{\partial f(t)}{\partial t}=-\frac{p}{2}\exp{(-T(p-2)t)}\int_0^t\,du f^{p-1}(t-u)f(u)\exp{(-T(p-2)u)} 
\label{eqas12}
\ee

with the initial condition $f(0)=1$. But the equation for the overlap
$Q(t,s)$ is more complicated and cannot be reduced to a
time-translational invariant solution. Writing
$Q(t,s)=q(t,s)\exp{(-T|t-s|)}$ (with $q(t,t)=Q_d(t)$ we obtain the following equations,

\bea
\frac{\partial q(t,s)}{\partial t}=\frac{p}{2}\int_0^s\
du\exp{(-2T(s-u))}\bigl (f(s-u)q^{p-1}(t,u)-f^{p-1}(t-u)q(s,u)\bigr )\\
-\frac{p}{2}\int_s^t\
 du f^{p-1}(t-u)q(u,s)\exp{(-T(p-2)(t-u))}\label{eqas13a}\\
\frac{1}{2}\frac{\partial Q_d(t)}{\partial t}+T(Q_d(t)-{\cal K}(Q_d))=\frac{p}{2}\int_0^t\
du\exp{(-2T(t-u))}\bigl (f(t-u)q^{p-1}(t,u)-f^{p-1}(t-u)q(t,u)\bigr )\label{eqas13b}~~~.
\eea

A  time-translational invariant solution for $q(t,s)$ does not exist for all
times (contrarily to what happens for $C$ and $R$) because the first
integral in the right hand side of eq. (\ref{eqas13a}) does not
vanish. 

Previous equations are solvable for $p=2$ and ${\cal K}=1$. In this case we
get,

\bea
\frac{\partial f(t)}{\partial t}=-\int_0^t\,du f(t-u)f(u)\label{eqas14a}\\
Q_d(t)=1-(1-Q_d(0))\exp{(-2Tt)}\label{eqas14b}\eea

and there is no DS transition for $p=2$ as expected (i.e. $T_0=0$). For
a generic cross-correlation ${\cal K}$ let us note that the 
equation (\ref{eqas13b}) reduces to equation (\ref{K1}) so there will be
different asymptotic values depending on the value of ${\cal K'}(Q^*)$ at the
different set of fixed points $Q^*={\cal K}(Q^*)$. It is notorious that
the case $p=2,\alpha=-1$ at finite temperature reduces to the infinite temperature
case for any $\alpha$. This is closely related to the fact that the
stationary solution in this case coincides with the random initial
configuration although this is not true anymore for $p>2$. For $p>2$ the lowest
$T_0$ temperature for maximal cross-correlation ${\cal K}=1$ becomes finite
and this is due to the rugged structure of the {\em force} landscape.

\begin{figure}
\begin{center}
\leavevmode
\epsfysize=160pt{\epsffile{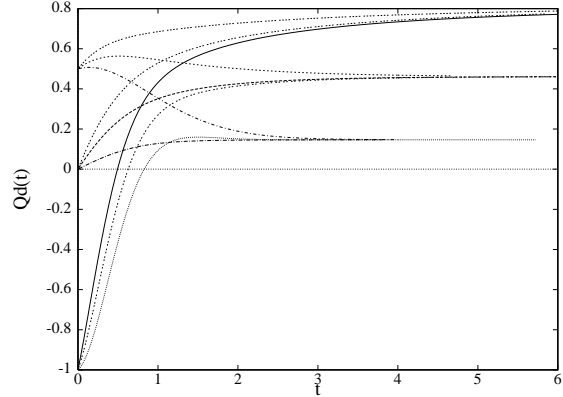}}
\end{center}
  \protect\caption[1]{$Q_d(t)$ ($p=3, \alpha=-1/2$, ${\cal K}=1$) with $Q_d(0)=0,0.5,-1$ as a
  function of time for temperatures $T=0.1,0.3,0.5$ (from bottom to
  top).\protect\label{FIGASC} }
\end{figure}

Equations for the damage for $p>3$ are difficult to solve. We have not
succeeded to obtain an analytical expression for the asymptotic values
as well as for $T_c$ and $D_{\infty}$. Numerical integrations of the
equations reveal that the transition persists at finite temperatures for
$p>2$. Figure~\ref{FIGASC} shows the overlap $Q_d(t)$ as a function of
time for different temperatures for $p=3$, ${\cal K}=1$. In this case the transition
is located at $T_0\simeq 0.595\pm 0.005$. Note that again relaxation to
the stationary state is faster than in the case $\alpha=1$ and
$D_{\infty}$ is again independent of $D(0)$.

\begin{figure}
\begin{center}
\leavevmode
\epsfysize=160pt{\epsffile{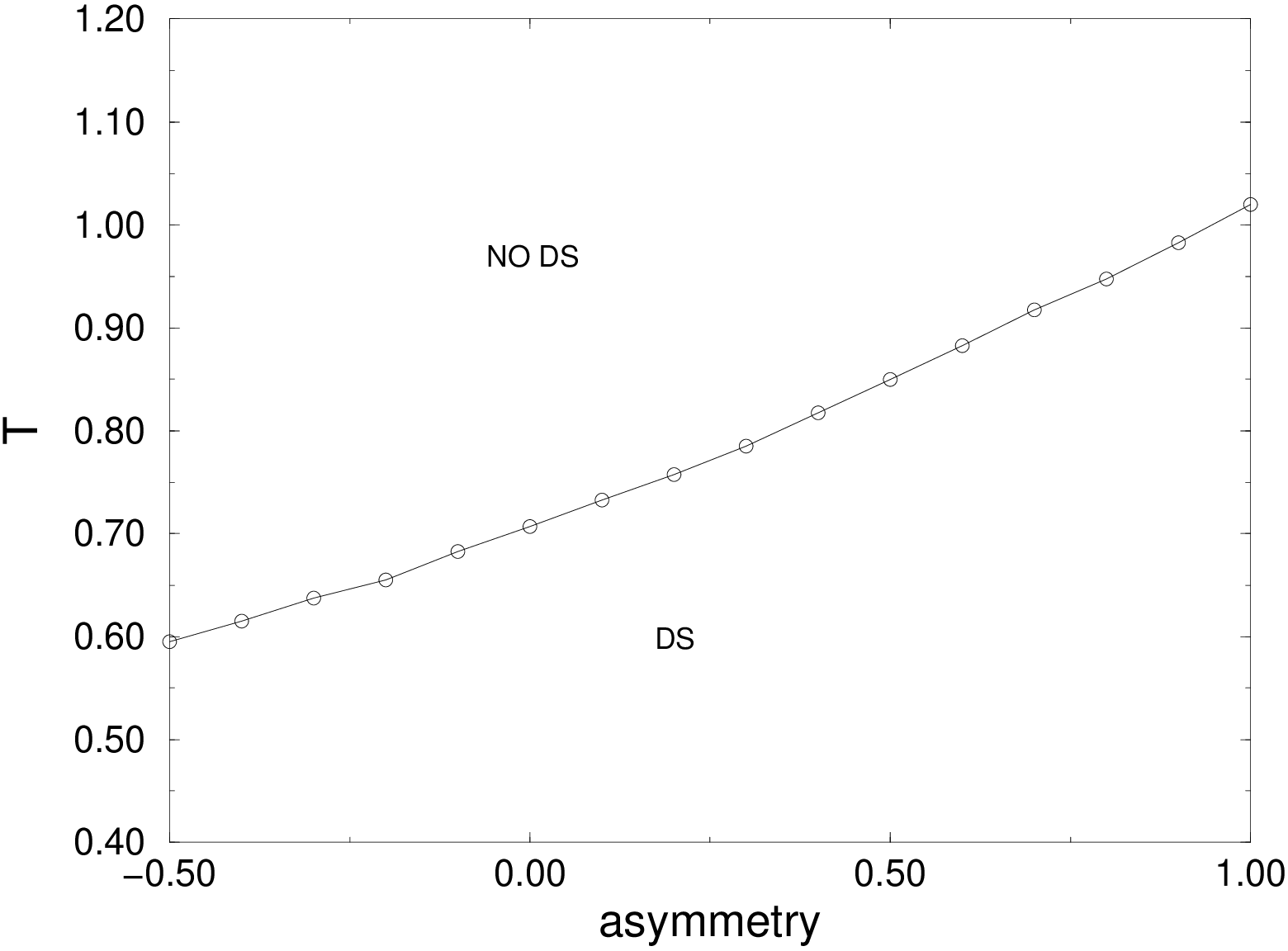}}
\end{center}
  \protect\caption[1]{$T-\alpha$ phase diagram for $p=3$, ${\cal K}=1$. The line is a guide to the eye.\protect\label{FIGASD} }
\end{figure}

These results are quite appealing. Here we find a DS transition in the
presence of a time-translational invariant solution for $C$ and $R$, i.e
when the system starts already in the stationary state. This is in
agreement with the results of previous subsection $III.B$ for $\alpha=1$
where a DS transition was found (at the same temperature than starting
from random initial conditions) when the system already started in the
stationary state. Figure~\ref{FIGASD} summarizes our results. We show
the $T-\alpha$ phase diagram of the DS transition for $p=3$.

Let us remark our final conclusion for this section. A DS transition is
present for all models with $p>2$ above the TAP temperature where an
exponentially large number of states appear. For a given choice of
${\cal K}$ such that $Q=1$ is stable at infinite temperature, this
transition has the following interesting properties: a) The asymptotic
distance is independent of the initial distance; b) Is also independent
of the type of initial configuration; c) Is stable in the presence of
asymmetry.  (but is unstable for $p=2$ in agreement with results derived
by Crisanti and Sompolinsky). This suggests that $D_{\infty}$ has some
of the crucial properties to be a good dynamical order parameter. The
correlator $Q(t,s)$ is not time-translational invariant in the time
scale in which the correlation and the response are. 

As we said previously we expect the properties of this transition to strongly
depend on the type of dynamics through the cross-correlator ${\cal
K}$. In the next section, we will discuss discrete (Monte Carlo)
dynamics, the case in which different algorithms correspond to 
different cross-correlators.

\section{Damage spreading in discrete glassy models}

Up to now we have considered the DS problem in case of a dynamics
continuous in time such as Langevin dynamics. 
Here we want to investigate damage spreading and in particular the
existence of $T_0$ for discrete dynamics such as Monte Carlo algorithms.

The analogue of the functions ${\cal K}$ for continuous dynamics are
the different algorithms used in Monte Carlo dynamics for discrete
dynamics. So different Monte Carlo algorithms determine different
types of correlations between the noises.
As for the cross-correlator ${\cal K}$ this implies that the
algorithms determine the structure of the high-temperature fixed
points. Their instabilities determine the subsequent low-temperature
behavior. 
As we have already commented in the
introduction we therefore do not expect the DS transition to be universal and
the results of this section aim to be compared with the results
already obtained for the Langevin dynamics in a continuous system.

One of the essential ingredients for the DS transition is the presence
of two competing effects: a synchronizing force (the stochastic noise
$\eta$ in the Langevin dynamics) and a landscape-dependent force which
pulls configurations apart one from each other into different
directions. In the case of a discrete (Monte Carlo) dynamics the equivalent role of the stochastic noise is played
by the set of random numbers generated during the Monte Carlo
updates. Now, the random number in the Monte Carlo algorithm
(uniformly chosen between 0 and 1) determines the probability of a
move depending also on the configuration of the system. This last
dependence corresponds to the role played by the cross-correlator
${\cal K}$ in the Langevin case where the two noises $\eta,\eta'$ may
be different depending on the value of ${\cal K}(C,C')$. Roughly
speaking the Metropolis algorithm for Monte Carlo dynamics corresponds
to the case ${\cal K}(Q)=Q$ for Langevin. It is easy to check
that, at infinite temperature, the fixed-points in both dynamics are
the same.

For continuous (Langevin) dynamics we had the freedom to choose the maximal
cross-correlation ${\cal K}=1$. 
For discrete dynamics, however, this is not the case.
There are several well known algorithms in the Monte Carlo approach
according to which updating rule they use. For instance, Metropolis,
Glauber, or heat-bath. Among these, the last one is the only one which
has a unique fixed point $Q^*=1$ at infinite temperature. So,
heat-bath dynamics is the closest case (but different) to the ${\cal
K}=1$ of Langevin dynamics. Here, our numerical investigation will
focus on this type of discrete dynamics. Let us note that the other
algorithms may show different behavior (due to the presence of other
infinite-temperature fixed points) and consequently also different DS
transitions. This non-universality of the DS transition (as in our previous analysis of the Langevin case) has received some attention in
the literature \cite{HR2}.

\subsection{Damage spreading in the Bernasconi model}

Here we will analyze the Monte Carlo dynamics with the heat-bath
algorithm for the Bernasconi model \cite{BERN}. This is a long-range
interaction model without disorder which is known to have a glassy
behavior being in the universality class of spin-glass models with one
step of replica symmetry breaking \cite{MPR}. Consequently, its
dynamical behavior is the same as predicted by the mode-coupling
theory.

The Bernasconi model (for simplicity we will consider the closed model,
see \cite{MPR} for more details) consists of $N$ Ising spins $\s_i=\pm 1$
in a one dimensional chain interacting through a long-range four-spin
interaction. It is defined by the following Hamiltonian, 

\be
H=\frac{1}{N}\sum_{k=1}^{N-1} C_k^2
\label{Berna}
\ee

with $C_k=\sum_{j=1}^N \s_j\s_{j+k}$ and where we take periodic boundary conditions $\s_i=\s_{i+N}$.

In this model there are particular values of $N$ for which the ground
state is exactly known \cite{MPR}. The interest of this model is that it
behaves like a disordered spin glass in the absence of explicit quenched
disorder in the Hamiltonian. Apparently, disorder is self-induced by the
dynamics \cite{KTH0,MPR,MPR2,BM}. This means that dynamics itself
generates slow evolving variables which effectively act as quenched
disordered fields.  This model is characterized by three temperatures, a
melting crystal-liquid first-order transformation temperature $T_{M}$, a
dynamical transition temperature $T_d\simeq 0.5$ \cite{MR} where the
relaxation time diverges and ergodicity breaks and, finally, a static
(or glass) transition temperature $T_s\simeq 0.25$ where replica symmetry breaks
and the configurational entropy collapses (this is the ideal glass transition
predicted in the AGM theory).

In the heat-bath algorithm, to go from a configuration $\{ \s_i(t)\}$,
at a given time $t$, to a configuration $\{ \s_i(t+\Delta t)\}$ at the
next time-step $t+\Delta t$ with $\Delta t=1/N$, a spin $\s_k$ is
chosen at random among the $N$ spins to be updated.  The probability to
put the spin up or down is decided according to the intensity of the
local field acting on that spin. More precisely, if we write the Hamiltonian
(\ref{Berna}) in terms of the local field $H=-\sum_k\,h_k\,\s_k$ then the
probability of putting the spin $\s_k$ up ($\s_k=1$) or down ($\s_k=-1$) at time
$t+\Delta t$ is given by,

\be P(\s_k(t+\Delta t)=\s)= \frac{1}{2}+\frac{1}{2}
\tanh{( \beta h_k(t)\s )}
\label{prob}
\ee 

with $\s=\pm 1$ and $h_k(t)$ is the local field acting on the spin $k$
at time $t$. Note that the probability (\ref{prob}) does only depend
on the local field acting on the spin $k$ and not on the actual value
of that spin at time $t$. Then, a random number $z(t)$ with a uniform
distribution between $0$ and $1$ can be introduced and spins are
sequentially updated according to the dynamical rule

\be \s_k(t+\Delta t)= 
{\rm sign}\left[\frac{1}{2}+\frac{1}{2} \tanh{( \beta
h_k(t))} -z(t)\right]
\label{heat-bath}
\ee

\begin{figure}
\begin{center}
\leavevmode
\epsfysize=150pt{\epsffile{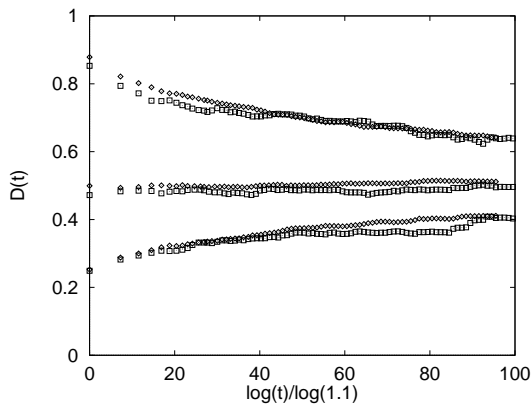}}
\end{center}
  \protect\caption[5]{The distance $D$ averaged per interval
    $(a^k,a^{k+1}-1)$ as a function of logarithmic
    time with base $a=1.1$ for temperature $T=0.3$.
The upper two curves are the result of initial condition $D(0)=1$ for
    $N=1000$ (boxes) and $N=5000$ (diamonds).
In the same manner, the middle (lower) curves are the result of
    $D(0)=0.5$ ($D(0)=0.1$).
\protect\label{FIGevolT.3allb} }
\end{figure}

With this rule (\ref{heat-bath}) we have studied numerically the damage
spreading of three different initial conditions, as in the $p$-spin
model: a) anti-correlated random
initial conditions with $D(0)=1$, b) uncorrelated random initial conditions with
$D(0)=0.5$ and c) partially correlated random initial conditions with
$D(0)=0.1$.  
For each of these cases, the distance $D(t)$ is computed up to
$100000$ and $10000$ Monte-Carlo time-steps for $N=1000$ and
$N=5000$ respectively.
To analyze the data, the logarithmic time with base $a=1.1$ is
considered.
Moreover, the data is averaged in intervals of the form ($a^k,a^{k+1}-1$)
with $k$ a positive integer.
For $T=0.3$ the evolution of $D(t)$ is plotted in figure~\ref{FIGevolT.3allb}.
  
\begin{figure}
\begin{center}
\leavevmode
\epsfysize=180pt{\epsffile{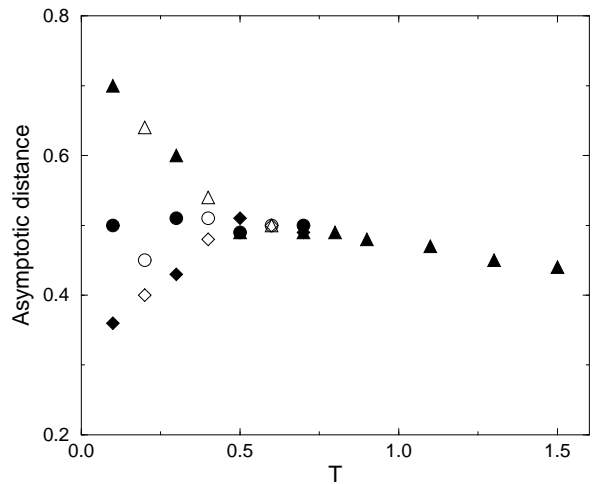}}
\end{center}
  \protect\caption[6]{The asymptotic distance $D_{\infty}$ as a
  function of temperature for three different initial conditions:
  $D(0)=1$ with $N=5000$($1000$) as closed(open) triangles,
  $D(0)=0.5$ with $N=5000$($1000$) as closed(open) circles and
  $D(0)=0.1$ with $N=5000$($1000$) as closed(open) diamonds.
\protect\label{FIGbern} }
\end{figure}

To obtain the asymptotic value $D_{\infty}$ from figures like
\ref{FIGevolT.3allb}, a power law fit for low temperatures is used.
For high temperatures an exponential fit up to $2000$ Monte-Carlo
steps in real time is used. 
The results are displayed in figure~\ref{FIGbern}.
One observes that the distance does not vanish for any finite
temperature.  Moreover, figure~\ref{FIGbern} indicates the existence of
a temperature $T_1$ above which the asymptotic distance is independent
of the initial distance.  Below this temperature $T_1$, $D_{\infty}$
does seem to be dependent of the initial distance. This dependence is
supported by a numerical extrapolation which could well fail when going
to enormously large time scales. Still, what we certainly find is the
appearance of a dynamical transition temperature $T_1\simeq 0.5$ in very
good agreement with the transition $T_d$ where ergodicity breaks
\cite{MPR,MR}.

\begin{figure}
\begin{center}
\leavevmode
\epsfysize=150pt{\epsffile{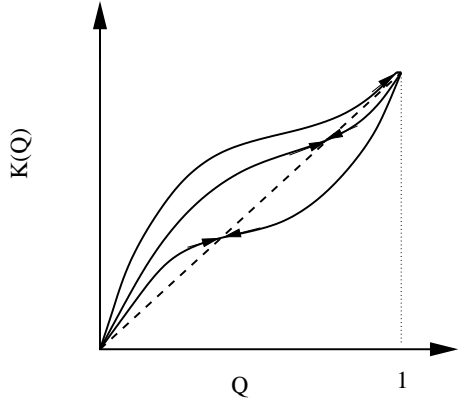}}
\end{center}
  \protect\caption[1]{Possible flow diagram at different temperatures for 
heat-bath dynamics\protect\label{FIXED3} }
\end{figure}

The behavior we find here, when compared to the previous Langevin
analysis for the $p$-spin model, may appear quite different. But a
careful analysis reveals that this is not the case. If we consider that
${\cal K}$, for the Langevin case, is a generic function which may
depend on the overlap as well as on the temperature $T=1/\beta$, we may
then imagine a situation such as depicted in figure~(\ref{FIXED3}) where
the infinite-temperature fixed-point $Q=1$ becomes unstable as soon as
$\beta$ is finite. In this case the asymptotic distance would be a non
trivial function of $\beta$ and the damage-spreading transition could
well happen at the usual dynamical transition $T_d$ where ergodicity is
broken. The dependence of the asymptotic overlap on the initial value
could be a consequence of the presence of different fixed points at low
temperatures.

In the most general case one could imagine a scenario with three
possible different regimes. A high-temperature regime $T>T_{0}$ where
$D_{\infty}=0$ independently of the initial distance $D_0$. An
intermediate regime $T_1<T<T_0$ where $D_{\infty}=D_{\infty}(T)$ is not
zero but independent of the initial distance (this regime would
correspond to the appearance of temperature dependent fixed point for
$\beta$ finite as depicted in figure~(\ref{FIXED3})). And finally a
low-temperature regime $T<T_1$ where $D_{\infty}=D_{\infty}(T,D_0)$
depends on both temperature and initial distance.  The results we find
for the Bernasconi model are the same as found by Derrida and Weisbuch
\cite{DW} for the Sherrington-Kirkpatrick model. The fact that in this
model $T_0\to\infty$ is related to the infinite-range character of the
interaction. Actually, Derrida\cite{DER} has found numerical evidence
that for finite-dimensional spin glasses there exists a range of
temperatures where the asymptotic distance vanishes and $T_0$ is finite.
The dependence of the asymptotic overlap on the initial condition found
here and in \cite{DER} below $T_1$ could well be an artifact of the
large-time extrapolation where the time window simulated and the size as
well are not sufficiently large. Unfortunately, it is not easy to
simulate very large times and sizes in infinite-ranged models like the
present one.

The results of this section show that, the DS transition is very close
(and probably coincides) with the dynamical transition temperature $T_d$
(below which the system never attains equilibrium and ergodicity is
broken). Nevertheless, in this case the asymptotic damage below $T_1$
apparently depends on the initial condition (and probably on the type of
initial condition as well) although such a firm conclusion needs more
understanding of damage-spreading transitions for generic updating rules.

\section{Conclusions}

In this paper we have studied the problem of damage spreading in the
mode-coupling theory of glasses. Mode-coupling theory is well known to
describe relaxation processes in glasses in the mean-field
approximation. A simple way to obtain the mode-coupling equations is
analytically solving the dynamics of multipsin interaction spherical
spin-glass models. These models are
characterized by the presence of a huge number of meta-stable states
(exponential with the system size) which appear at a temperature
$T_{TAP}$ higher than $T_d$ (where ergodicity breaks) and $T_s$ (where
replica symmetry breaks). Because
the phase space in this class of models is characterized by a extremely
rugged and complex free energy landscape they are good candidates to
study the landscape properties using techniques taken from dynamical
systems.

A very interesting technique which is able to probe the topological
features of the phase space landscape is damage spreading. This consists
in the study of the distance between the configurations of two
stochastic systems submitted to the same realization of the stochastic
noise but differing in the initial conditions. By the same realization
of the stochastic noise we mean noises that are statistically identical
although generally correlated through a function ${\cal K}(Q)$ which
satisfies the condition ${\cal K}(Q)\le {\cal K}(Q=1)=1$. 
In general, any choice for the correlator ${\cal K}$ alters the
results.
For Langevin dynamics, we have shown
that interesting results appear for the case ${\cal K}=1$. In
that case, both noises are identical for the two copies independently of
their configurations. This yields the lowest damage spreading
transition for which the $Q^*=1$ fixed-point is stable at high-temperatures.
Depending on the value of ${\cal K}'(Q=1)$ one finds a different damage
spreading transition temperature up to the limiting case ${\cal K}'(Q=1)=1$ 
(see eq.(\ref{equpper})) where $T_0=\infty$ and the fixed-point $Q^*=1$ becomes unstable.
Whether this holds for other type of dynamics is not studied and
remains unclear.

An exhaustive study has been done for the case ${\cal K}=1$ although
similar results are obtained for any ${\cal K}$ for which the solution
$Q=1$ is asymptotically stable. In this case, through functional methods
and using the Pade series expansion method (to make safe extrapolations
in the asymptotic long-time limit), we have shown the existence of a
damage spreading transition $T_0$ in general mode-coupling equations with any
degree of asymmetry in the interactions. In particular, in case of
symmetric interactions (where detailed balance holds) we have found
evidence for a damage spreading transition at a finite temperature. This
transition occurs at temperatures $T_0$ higher than $T_{TAP}$, this last
one being the temperature where an exponentially large number of
meta-stable states start to appear. The transition is characterized by a
dynamical order parameter $D_{\infty}$ which is the asymptotic distance
between the two evolving replicas. Interestingly, $D_{\infty}$ has the
good properties of order parameters being able to distinguish different
dynamical phases (in our case, there are two possible phases depending
whether $D_{\infty}$ vanishes or not).  These properties are: a)
$D_{\infty}$ is independent on the initial distance $D_0$ for a given
class of initial conditions, b) $D_{\infty}$ does not depend also on the
class of initial condition (whether they are random or thermalized) c)
The DS transition is stable against the inclusion of asymmetry in the
interactions (i.e. against the violation of detailed balance in the
dynamics). Furthermore, regarding the mode-coupling equations with
asymmetry we have obtained some exact results for the asymmetric case
$\alpha=0$ and exactly solved the correlation and the response function
for the antisymmetric case $\alpha=-\frac{1}{p-1}$ which interestingly
turns out to be time-translational invariant. The existence of DS
transition in this case reveals that this transition already appears
when the system is time-translational invariant.

We insist on the fact that the precise value of the damage spreading
temperature $T_0$ as well as the asymptotic distance $D_{\infty}$ both
depend on the correlator ${\cal K}$ considered. This fact expresses
the non-universal character of this transition where the
cross-correlator ${\cal K}$ plays the equivalent role of a stochastic
noise for the dynamical order parameter $Q_d(t)$. Different functions
${\cal K}$ imply different dynamical phase transitions so their
physical significance must be appropriately interpreted. In this
direction we have tried to interpret our results in terms of saddle
points in phase space for those cases where $Q^*=1$ is an
asymptotically stable solution at high temperatures. The essentials of
the argument where given in section II where it was shown that for
close enough initial conditions an instability in the $Q=1$
high-temperature solution is due to the presence of vanishing modes in
the Hessian of the potential function. This signals the presence of
saddle points in the free energy landscape, of which an infinite
amount yields a DS transition.
This certainly happens
at $T_{TAP}$ but we have found that the transition occurs already at a
temperature $T_0$ much above that temperature. How can we reconcile
our results with that?  At $T_{TAP}$ the number of meta-stable states
is exponentially large with $N$. Probably for the DS transition to
occur it is only necessary that this number be a big (for instance , a
power of $N$) but not exponentially large. Consequently, a divergent
number of meta-stable states and saddle points (but not exponentially
large with $N$) should be enough to make the DS transition
appear. Unfortunately, the analytical computation of the temperature
$T_0$ by counting the number of meta-stable states is not so direct
because such a calculation involves the estimate of finite-size
corrections to the dominant saddle point calculation
\cite{BRMO}. Actually, the evaluation of finite-size corrections in
spin glasses (even in mean-field) is known to be quite difficult.
Concerning the role of the cross-correlations between the noises we
remind that the appearance of vanishing modes tends to be suppressed
by cross-correlations ${\cal K}<1$ which play the role (according to
eq.(\ref{new3})) of a thermal noise on the system described by the
effective distance variables $z_i=x_i-y_i$. This explains why ${\cal
K}=1$ yields the lowest damage spreading temperature. We must stress
that this transition should persist beyond mean-field. Actually the
presence of a huge number of saddle points in phase space is not an
exclusive mean-field feature but should persist in the presence of
short-range corrections. Short-range corrections dramatically modify
the height of the barriers but not their number which still could
remain as large as in the mean-field approximation.

To remark the advantage of considering Langevin dynamics for DS we have
also investigated the DS in discrete glassy models without disorder such
as the Bernasconi model with heat-bath dynamics. The difference between
any discrete dynamics and Langevin dynamics is that the type of
correlator ${\cal K}$ in the last case may be chosen at will. So the
case ${\cal K}=1$ which has been studied for Langevin dynamics cannot be
implemented for Monte Carlo dynamics. Each algorithm for Monte Carlo
dynamics defines a given ${\cal K}$ so the study of damage spreading in
those cases remains more speculative because ${\cal K}$ is essentially
unknown. For the particular case of the heat-bath in the Bernasconi
model we find a DS transition which separates two regions, a region
$T>T_1$ where $D_{\infty}$ is independent of the initial distance but
finite and a region $T<T_1$ where $D_{\infty}$ depends on the initial
distance. $T_1$ coincides (within numerical precision) with the
mode-coupling transition $T_d$.  Although for heat-bath dynamics
$D_{\infty}$ depends on the initial distance we may not exclude that
convergence to the asymptotic limit is extremely slow and convergence
poor.

We end up our discussion describing some opening problems. Much work on
the DS has been devoted to the study of discrete dynamics in discrete
systems the situation being the opposite for continuous-time
dynamics. In this work we have shown that for Langevin dynamics we may
choose specific noise correlations so it is easier to interpret what
physical properties of the system we are looking at. It is difficult to
adscribe any physical significance to properties measured for arbitrary
discrete algorithms and this has been one of the major problems to
interpret the large amount of numerical results obtained in the study of
damage spreading in Monte Carlo simulations. Here we have seen that for
Langevin dynamics such a task turns out to be easier.  Still one of the
major tasks which remain open is to understand better under which
conditions there is a unique absorbing state for the damage. In another
direction, one would like to get analytical proofs and equilibrium based
analytical methods to investigate the connection between the DS
transition and the topological properties of the phase space (such as
the presence of saddle points in phase space). Finally, we would like to
extend our research to real structural glasses where it could be
interesting to study DS and investigate in which conditions such a
transition is a precursor of the glass transition.

{\bf Acknowledgments}.  We acknowledge stimulating discussions with
M. Campellone, S. Franz, W. A. Van Leeuwen, Th. M. Nieuwenhuizen and
M. Sellito.  We are grateful to T. Alarcon for a careful reading of the
manuscript and Ian Campbell for pointing us some mistakes in the
original version.  We also thank an anonymous referee for noting the
crucial role of the correlator ${\cal K}$ in DS transitions. F.R acknowledges
funding from the Spanish government through the project PB97-0971.

\appendix

\section{}

In this appendix we describe some of the main steps necessary to derive
equations (\ref{eqa5a})-(\ref{eqa6b}). Consider two replicas
$\lbrace\s\rbrace,\lbrace\tau\rbrace$ submitted to the dynamics (\ref{eqa1}),

\be
\frac{\partial \s_i}{\partial t}=F_i(\lbrace\s\rbrace)-\mu\s_i+\eta_i
\, , \qquad 
\frac{\partial \tau_i}{\partial t}=F_i(\lbrace\tau\rbrace)-\mu\tau_i+\eta'_i
\label{apa}
\ee

\noindent
where $F_i$ is the force acting on the spin $\s_i$ due to the
interaction with the rest of the spins eq. (\ref{eqa2}). The noises
$\eta$,$\eta'$ are generically correlated $<\eta_i(t)\eta'_j(s)>=2T{\cal
K}(\lbrace\s\rbrace,\lbrace\tau\rbrace)\delta(t-s)\delta_{ij}$. Following
Domany and Hinrichsen \cite{HR2} we will suppose that the correlator
${\cal K}$ is a generic function of the overlap due to the
infinite-ranged nature of the model. So we will take ${\cal
K}(\lbrace\s\rbrace,\lbrace\tau\rbrace)$ a generic function of the
equal-times overlap ${\cal K}(Q_d)$ where
$Q_d=\frac{1}{N}\sum_{i=1}^N\s_i\tau_i$. Following the same steps as is
usually done in the study of the dynamics of a single replica (see
\cite{BARRAT} for details) we may write the generating functional for
the dynamics in the Ito prescription,

\be {\cal
Z}_{dyn}=\int[d\s d\tau]\delta(\dot{\s}_i-F_i(\s)+\mu\s_i-\eta_i)
\delta(\dot{\tau}_i-F_i(\tau)+\mu\tau_i-\eta'_i)\label{apa3}
\ee

Introducing a new set of fields $\hat{\s}_i$,$\hat{\tau}_i$ and
averaging over the noise we get,

\be
{\cal
Z}_{dyn}=\int[d\s d\hat{\s} d\tau d\hat{\tau}]\exp\bigl ( {\cal
L}(\s,\hat{\s},\tau,\hat{\tau})\bigr )
\label{apa4}
\ee

where

\bea
{\cal L}(\s,\hat{\s},\tau,\hat{\tau})=-i\int dt \sum_i
\hat{\s}_i^t(\dot{\s}_i^t-F_i(\s^t)+\mu\s_i^t)-i\int dt \sum_i
\hat{\tau}_i^t(\dot{\tau}_i^t-F_i(\tau^t)+\mu\tau_i^t) -\nonumber\\T\int dt 
\sum_i\bigl ( \frac{1+{\cal K}}{2}(\hat{\s}_i^t+\hat{\tau}_i^t)^2
+\frac{1-{\cal K}}{2}(\hat{\s}_i^t-\hat{\tau}_i^t)^2\bigr )
\label{apa5}
\eea

Because $Z_{dyn}=1$ we may average the dynamical partition function over
the disorder. We use the cumulant expansion and retain only the
first and second order terms $ \overline{\exp(V)}\simeq exp\bigl ( \overline{V}+\frac{1}{2}(\overline{V^2}-(\overline{V})^2)\bigr)$
where $\overline{(.)}$ stands for disorder average. Using
eq. (\ref{eqcorrF}) we obtain the
final result,

\be
\overline{{\cal Z}_{dyn}}=\int [d\s d\hat{\s} d\tau
d\hat{\tau}]\exp\bigl ({\cal S}(\s,\hat{\s},\tau,\hat{\tau})\bigr )
\label{apa7}
\ee

where

\bea
{\cal S}(\s,\hat{\s},\tau,\hat{\tau})=-i\int dt \sum_i
\hat{\s}_i^t(\dot{\s}_i^t+\mu\s_i^t)-i\int dt \sum_i
\hat{\tau}_i^t(\dot{\tau}_i^t+\mu\tau_i^t) -T\int dt
\sum_i\bigl ( \frac{1+{\cal K}}{2}(\hat{\s}_i^t+\hat{\tau}_i^t)^2
+\frac{1-{\cal K}}{2}(\hat{\s}_i^t-\hat{\tau}_i^t)^2\bigr )
\nonumber\\
-\frac{1}{2}\int\ dt\ ds \Bigl ( \sum_i\hat{\s}_i^t\hat{\s}_i^s f'(C(t,s))+
\sum_i\hat{\tau}_i^t\hat{\tau}_i^s f'(D(t,s))+\sum_i\hat{\s}_i^t\hat{\tau}_i^s f'(Q(t,s))+
\sum_i\hat{\tau}_i^t\hat{\s}_i^s f'(Q(s,t)) \Bigr )\nonumber \\
-\frac{\alpha}{2N}\int\ dt\ ds \sum_{ij}\Bigl
(\hat{\s}_i^t\hat{\s}_j^s\s_i^s\s_j^t\ f''(C(t,s))+
\hat{\tau}_i^t\hat{\tau}_j^s\tau_i^s\tau_j^t\ f''(D(t,s))+
\hat{\s}_i^t\hat{\tau}_j^s\tau_i^s\s_j^t\  f''(Q(t,s))+
\hat{\tau}_i^t\hat{\s}_j^s\s_i^s\tau_j^t\  f''(Q(s,t)) \Bigr )
\label{apa8}
\eea

The correlation functions $C,D,Q$ are defined by,

\be
C(t,s)=\frac{1}{N}\sum_i<\s_i^t\s_i^s>\label{apa9} \, , \quad
D(t,s)=\frac{1}{N}\sum_i<\tau_i^t\tau_i^s>\label{apa10} \, , \quad
Q(t,s)=\frac{1}{N}\sum_i<\s_i^t\tau_i^s>\label{apa11}
\ee

After introducing the necessary response functions, for instance 
$R(t,s)=i<\hat{\s}_i^s\s_i^t>$ we may solve the saddle point
equations. Introducing also appropriate causality relations (for instance
$<\hat{\tau}_i^s\s_i^t>=0$) we obtain the desired set of
equations. Note that only the dynamical equation for the equal times
overlap $Q_d(t)=Q(t,t)$ depends on the correlator ${\cal K}$.

\section{}

In this appendix we prove the existence of the lower and upper bounds 
eq. (\ref{equpper}) for $T_0$ in case $\alpha=1$, ${\cal K}=1$. 

\noindent
To obtain the lower bound we start from eq. (\ref{eqa6b}) and write,

\be
\frac{1}{2}\frac{\partial Q_d(t)}{\partial t}+\mu(t)Q_d(t)-T{\cal K}(Q_d)-
\frac{p}{2}\int_0^t du R(t,u)Q^{p-1}(t,u)
-\frac{p(p-1)}{2}
\int_0^t du R(t,u)Q(t,u)C^{p-2}(t,u)=0\label{apb1}~~~~~~.
\ee

Let us assume now that we are in the high-temperature phase
where the $R$ and $C$ are time-translational invariant and the
fluctuation-dissipation theorem is satisfied,

\be 
TR(t-s)= \frac{\partial C(t-s)}{\partial s}~~~~.
\label{apb2}
\ee

At first glance, this condition may seem too strong. Nevertheless,
assuming FDT is allowed if we are in the high-temperature phase. At the
DS temperature (which is deep in the high temperature phase, as we show
in detail in sections III.A and III.B) the typical relaxation time for
the quantities $C$ and $R$ is finite so we may introduce FDT with no
harm.  On the other hand, section III.B shows that the DS transition is
already present if we start from an equilibrium condition where
condition (\ref{apb2}) is satisfied for all times.

Using the inequality $Q(t,s)\le Q_d(s)$ where $t>s$ we obtain,

\be
\frac{1}{2}\frac{\partial Q_d(t)}{\partial t}+\mu(t)Q_d(t)-T{\cal K}(Q_d)-
\frac{\beta}{2}\int_0^t du \frac{\partial C^p(t-u)}{\partial u}Q_d^{p-1}(u)
-\frac{\beta(p-1)}{2}
\int_0^t du \frac{\partial C^p(t-u)}{\partial u}
Q_d(u)\le 0\label{apb3}~~~~~~.
\ee

Doing an integration by parts for the two integrals appearing in
(\ref{apb3}) we may write,

\bea
\frac{1}{2}\frac{\partial Q_d(t)}{\partial t}+\mu(t)Q_d(t)-T{\cal K}(Q_d)-
\frac{\beta}{2}\bigl (Q_d^{p-1}(t)-Q_d^{p-1}(0)C^p(t)-
\int_0^t du \frac{\partial Q_d^{p-1}(u)}{\partial u}C^p(t-u)\bigr )\nonumber\\
-\frac{\beta(p-1)}{2}\bigl ( Q_d(t)-Q_d(0)C^p(t)-
\int_0^t du \frac{\partial Q_d(u)}{\partial u}C^p(t-u) \bigr
)~~~\le 0.\label{apb3b}
\eea

Noting that $Q_d(t)$ is a monotonous increasing
function of time (again, this is true only above the DS transition) we may
write,
\be
\int_0^t du \frac{\partial Q_d^{p-1}(u)}{\partial u}C^p(t-u)\ge
0\label{apb4} \, , \qquad
\int_0^t du \frac{\partial Q_d(u)}{\partial u}C^p(t-u)\ge 0\label{apb5}
\ee

so, in the long-time limit $t\to\infty$ where the $C(t)$ vanishes we get,

\be
\frac{1}{2}\frac{\partial Q_d(t)}{\partial t}+T(Q_d(t)-{\cal K}(Q_d))-
\frac{\beta}{2} (Q_d^{p-1}(t)-Q_d(t))\le 0
\label{apb6}
\ee

where we have replaced $\mu(t)$ by its equilibrium value
$T+\frac{p\beta}{2}$. Finally, due to the monotonicity of $Q_d(t)$ we
reach the desired inequality,

\be
0\le \frac{1}{2}\frac{\partial Q_d(t)}{\partial t}\le T({\cal K}(Q_d)-Q_d(t))+
\frac{\beta Q_d(t)}{2} (Q_d^{p-2}(t)-1)
\label{apb7}
\ee

If we approach the DS transition $T_0$ from above we expect $Q_d(t)$ to
relax very slowly to its asymptotic value $Q_d(\infty)=1$. So, we expect
$|\frac{\partial^2 Q_d(t)}{\partial t^2}| \le |\frac{\partial
Q_d(t)}{\partial t}| \le (1-Q_d(t))$ for large times. We may
differentiate the inequality (\ref{apb7}) and put $Q_d=1$ which yields

\be
0\ge \frac{1}{2}\frac{\partial^2 Q_d(t)}{\partial t^2}\ge 
\frac{\partial Q_d(t)}{\partial t} (({\cal K}'(1)-1)T+\frac{\beta(p-2)}{2})
\label{apb8}
\ee

At the DS transition the inequality is satisfied only if $T_0\ge
\sqrt{\frac{p-2}{2(1-{\cal K'}(1))}}$. 
For case $p=2$ the second member in the last expression of the
inequality (\ref{apb8}) vanishes so the inequalities are never violated
as soon as $Q_d\le 1$. This means, that there is no DS transition in the
high-temperature phase for $p=2$. The only possible transition occurs
when $C(t,s)$ and $R(t,s)$ are no time-translational invariant anymore
and this may happen only at $T_d$. Actually, calculations of Stariolo
\cite{STA} show that the DS transition is present at the static
transition temperature $T_s$ (which is equal to $T_d$).

As a curiosity (for which we have no analytical derivation) we note that
this lower bound (for the case ${\cal K}=1$) seems to coincide with the
exact DS transition temperature for the fully asymmetric case $\alpha=0$
(see section III.C).

On the other hand, the upper bound can be obtained via a linear
stability analysis of the equations (\ref{eqa5c}),(\ref{eqa6b}) around
the HT solution.
Writing $Q_d(t)=1-\eps f(t)$ and $Q(t,s)=C(t-s)-\eps g(t,s)$ with
$f(0)=1$ and $g(t,t)=f(t)$ yields for equation (\ref{eqa6b}) in the
large $t$ limit, 

\be
\frac{1}{2}\frac{\partial f}{\partial t}=-(T(1-{\cal K'}(1))-\frac{p\beta}{2})f-
\beta p \int_0^tdu C^{p-1}(t-u)\frac{\partial g(t,u)}{\partial u}~~~.
\label{eqf}
\ee

\noindent Finally, the inequality $\frac{\partial g(t,u)}{\partial
u}\ge 0$ yields the upper bound.

\hspace{-2cm}

\vfill

\end{document}